\documentclass[twocolumn,tighten,times]{aastex61}
\usepackage{color}
\usepackage{graphicx} \usepackage{epstopdf}
\usepackage{natbib}
\usepackage{longtable}

\newcommand\Wprime{W$^{\prime}$}
\newcommand\wprime{\Wprime}
\newcommand\Ttype{\ensuremath{T_\mathrm{type}}}
\newcommand\Reff{\ensuremath{R_e}}
\newcommand{\lsim}{\ \raise -2.truept\hbox{\rlap{\hbox{$\sim$}}\raise 5.truept\hbox{$<$}\ }} 
\newcommand{\gsim}{\ \raise -2.truept\hbox{\rlap{\hbox{$\sim$}}\raise 5.truept\hbox{$>$}\ }}
\newcommand\ustar{\ensuremath{u^*}}
\newcommand\gi{\ensuremath{(g{-}i)}}
\newcommand\gz{\ensuremath{(g{-}z)}}
\newcommand\ui{\ensuremath{(u^{*}{-}i)}}
\newcommand\ug{\ensuremath{(u^{*}{-}g)}}
\newcommand\uz{\ensuremath{(u^{*}{-}z)}}
\newcommand\iz{\ensuremath{(i{-}z)}}
\newcommand{\feh}{$\mathrm{[Fe/H]}$\ }
\newcommand\kms{km~s$^{-1}$}

\newcommand\mM{\ensuremath{(m{-}M)}}
\newcommand\mMacs{\ensuremath{(m{-}M)_\mathrm{ACS}}}
\newcommand\gzacsvcs{\ensuremath{(g{-}z)_\mathrm{ACS}}}
\newcommand\sna{SNe\,Ia}
\newcommand\hst{{\it HST}}

\newcommand\vi{{\ifmmode{(V{-}I)}\else$(V{-}I)$\fi}}
\newcommand\gI{{\ifmmode{(g{-}I)}\else$(g{-}I)$\fi}}
\newcommand\gIacs{{\ifmmode{(g_{475}{-}I_{814})}\else$(g_{475}{-}I_{814})$\fi}}

\newcommand\Mbar{\ensuremath{\overline{M}}}
\newcommand\Mbari{\ensuremath{\overline{M}_i}}

\newcommand\mbar{\ensuremath{\overline{m}}}
\newcommand\mbari{\ensuremath{\overline{m}_i}}

\newcommand\Mibar{\ensuremath{\overline{M}_i}}

\newcommand\mibar{\ensuremath{\overline{m}_i}}

\newcommand\lta{\lesssim}
\newcommand\gta{\gtrsim}
\newcommand\rmsmad{\ensuremath{\mathrm{rms}_\mathrm{MAD}}}

\makeatletter
  \newcommand{\miniscule}{\@setfontsize\miniscule{2}{4}}
\makeatother

\submitjournal{ApJ}

\shorttitle{SBF Calibration for the NGVS}
\shortauthors{Cantiello et al.}

\begin{document}


\title{The Next Generation Virgo Cluster Survey (NGVS).\ XVIII.~Measurement
  and Calibration of Surface Brightness Fluctuation Distances for Bright Galaxies in Virgo (and Beyond)}

\author[0000-0003-2072-384X]{Michele Cantiello}
\affil{INAF Osservatorio Astronomico d'Abruzzo, via Maggini, snc, 64100, Italy}
\email{cantiello@oa-teramo.inaf.it}
\author[0000-0002-5213-3548]{John P.\ Blakeslee}
\affil{National Research Council of Canada, Herzberg Astronomy and Astrophysics Research Centre, Victoria, BC, Canada}
\author{Laura Ferrarese}
\affil{National Research Council of Canada, Herzberg Astronomy and Astrophysics Research Centre, Victoria, BC, Canada}
\affil{Gemini Observatory, Northern Operations Center, 670 N.~A'ohoku Place, Hilo, HI 96720, USA}
\author{Patrick C\^ot\'e}
\affil{National Research Council of Canada, Herzberg Astronomy and Astrophysics Research Centre, Victoria, BC, Canada}
\author{Joel C.\ Roediger}
\affil{National Research Council of Canada, Herzberg Astronomy and Astrophysics Research Centre, Victoria, BC, Canada}
\author{Gabriella Raimondo}
\affil{INAF Osservatorio Astronomico d'Abruzzo, via Maggini, snc, 64100, Italy}
\author{Eric W.~Peng}
\affil{Department of Astronomy, Peking University, Beijing 100871, China}
\affil{Kavli Institute for Astronomy and Astrophysics, Peking University, Beijing 100871, China}
\author{Stephen Gwyn}
\affil{National Research Council of Canada, Herzberg Astronomy and Astrophysics Research Centre, Victoria, BC, Canada}
\author{Patrick R. Durrell}
\affil{Department of Physics and Astronomy, Youngstown State University, Youngstown, OH 44555, USA}
\author{Jean-Charles Cuillandre}
\affil{AIM Paris Saclay, CNRS/INSU, CEA/Irfu, Universit{\'e} Paris Diderot, Orme des Merisiers, France}

\begin{abstract}\noindent

We describe a program to measure surface brightness fluctuation (SBF)
distances to galaxies observed in the Next Generation Virgo Cluster
Survey (NGVS), a photometric imaging survey covering $104deg^2$ of the
Virgo cluster in the $u^{*},g,i,z$ bandpasses with the Canada-France
Hawaii Telescope.  We describe the selection of the sample galaxies,
the procedures for measuring the apparent $i$-band SBF magnitude
\mibar, and the calibration of the absolute Mibar as a function of
observed stellar population properties.  The multi-band NGVS data set
provides multiple options for calibrating the SBF distances, and we
explore various calibrations involving individual color indices as
well as combinations of two different colors. Within the color range
of the present sample, the two-color calibrations do not significantly
improve the scatter with respect to wide-baseline, single-color
calibrations involving $u^{*}$. We adopt the \uz\ calibration as
reference for the present galaxy sample, with an observed scatter of
0.11~mag. For a few cases that lack good ${u}^*$ photometry, we use an
alternative relation based on a combination of \gi\ and \gz\ colors,
with only a slightly larger observed scatter of 0.12~mag. The agreement
of our measurements with the best existing distance estimates provides
confidence that our measurements are accurate. We present a
preliminary catalog of distances for 89 galaxies brighter than
$B_T\approx13.0$ mag within the survey footprint, including members of
the background M and W Clouds at roughly twice the distance of the
main body of the Virgo cluster. The extension of the present work to
fainter and bluer galaxies is in progress.
\end{abstract}


\section{Introduction}

Reliable distance estimation is an essential prerequisite for
knowledge of the fundamental characteristics of structures in the
universe, including their size, luminosity, and mass.  Measurement of
significant samples of high-quality galaxy distances can enable
mapping of large-scale structures and velocity fields in the nearby
universe, where the peculiar velocity is often a substantial fraction
of the redshift.
Resolving the 3-D structure of galaxies within clusters requires
measurement precision for individual galaxies better than the ratio of
the cluster depth to its mean distance, or errors $<\,$0.1~mag even
for Virgo and Fornax, the only galaxy clusters within
$\approx\,$20~Mpc \citep[][hereafter B09]{mei07xiii,blake09}.

There are very few extragalactic distance indicators that are both
capable of this level of precision, and able to be applied to a large
fraction of galaxies within a given environment.  For instance, the
extensive review by \citet[][]{freedman10} discusses ``six
high-precision [extragalactic] distance-determination methods,''
namely, Cepheid variables, the tip of the red giant branch (TRGB),
water masers, the Tully-Fisher (TF) relation, Type~Ia supernovae
(\sna), and surface brightness fluctuations (SBF).  The first three of
these methods require high spatial resolution and have mainly been
used for nearby galaxies, i.e., at distances $\lesssim\,$20~Mpc
where peculiar velocities are comparable to the Hubble velocity; they
have been used to calibrate the latter three methods, which have
ranges extending out into the Hubble flow.

Most of the precision methods reviewed by \citet{freedman10} have been
applied to measure distances of galaxies in the Virgo cluster, thanks
to its proximity, with the exception being the water masers, which are
very rare.  However, not all of the methods are practical for mapping
the 3-D structure of the cluster.
For instance, Cepheids bolster the rung of the ladder by which our knowledge of
distances ascends from the solar neighborhood to the realm of the galaxies. However,
at extragalactic distances, Cepheids require many epochs of deep, high-resolution
observations and are only found within star-forming galaxies, not in the early-type
galaxies that populate dense environments.
The TF relation has traditionally produced distance errors $\gta\,$0.3~mag, much
larger than the depth of the Virgo cluster, but as \citet{freedman10} discuss, the
precision is vastly improved when using 3.6$\,\mu m$ photometry.  However, again the
method works best for spiral galaxies, rather than cluster ellipticals.
On the other hand, Type~Ia supernovae (\sna) occur in all galaxy types 
and are highly luminous, making them easily observable in the nearby universe.
However, they are rare events; according to the 
NASA/IPAC Extragalactic Database (NED),\footnote{{\url{http://ned.ipac.caltech.edu}}}
the last \sna\ to occur in the Virgo cluster cD galaxy M87 was a century ago,
SN\,1919A, and no confirmed \sna\ have occurred in the brightest cluster galaxy
M49, despite its enormous stellar mass.
The TRGB method is also applicable to all types of galaxies, but very deep
high-resolution imaging is required to reach a sufficient depth along the RGB in external
galaxies; consequently, very few galaxies in the Virgo cluster have TRGB distance
estimates \citep{durrell07,bird10,leejang17}.
%

Other methods with the potential for mapping the Virgo cluster region
include Mira variables \citep[e.g.][]{whitelock08} and the globular
cluster luminosity function \citep[GCLF; e.g.,][]{harris91}. Miras are
luminous asymptotic giant branch stars, especially bright in the
infrared, and likely present in most galaxies.  However, like the
other resolved stellar photometry methods, Miras require very high
spatial resolution and much longer temporal baselines then
Cepheids. Thus far, this method has only been used to distances of a
few Mpc \citep[]{rejkuba04}, but it may become much more far-reaching
with the combination of \textit{Gaia} and the \textit{James Webb Space
  Telescope} (\textit{JWST}). Gaia is expected to measure parallaxes
for $\sim40,000$ Miras \citep{robin12}, and \textit{JWST} will operate
in the near- and mid-IR regime, where Miras reach magnitudes brighter
than $\approx-8$~mag \citep[][$\approx-9.2$~mag at 8 $\mu
  m$]{feast14}.  GCLF measurements have been made for large samples of
galaxies in the Virgo and Fornax clusters \citep{jordan06,villegas10},
but in addition to the established trends with host galaxy luminosity,
the distance indicator likely depends in complex ways on galaxy
environment and dynamical history \citep[e.g.,][]{rejkuba12}. Overall,
it appears to be a less precise method than the others mentioned
above, especially for galaxies with small or modest GC populations.

Among the available extragalactic distance indicators, the SBF method, introduced by
\citet{ts88}, is the only distance indicator with high enough precision and wide enough
applicability to have clearly resolved the depth of the Virgo cluster
\citep{west00,mei07xiii} and detected the depth of the more compact Fornax cluster
(B09). This does not mean it is the most precise indicator, only that it is the best
suited for this particular problem. The reason is that, unlike Cepheids or the
TRGB, the SBF method does not require resolved stellar photometry, but like those methods,
it relies on well-understood stellar physics and can be calibrated from stellar
population observables, rather than assuming an empirical scaling law or a supposed
universal luminosity function.  In addition, unlike the \sna\ or maser methods, it is
based on a phenomenon present in all galaxies \citep[though with a predictability
  depending on stellar population; see the review by][]{blake12b}.

For early-type galaxies, or the relatively ``clean'' regions of
spirals without significant dust or recent star formation, if the data
are of sufficient depth for the SBF signal to exceed the photometric
noise, then the limiting factor is the ability to detect and remove
faint sources, especially globular clusters (GCs), that contaminate
the power spectrum of the fluctuations.  Excellent seeing and high
signal-to-noise greatly facilitate the rejection of such contaminants,
and therefore improve the precision of the method.  The SBF work by
\citet{mei07xiii} on the 3-D structure of Virgo used data of exquisite
quality from the ACS Virgo Cluster Survey \citep[ACSVCS,][]{cote04}, a
Large Program with the \textit{Hubble Space Telescope}
(\hst). However, as ground\-breaking as that survey was, its 100 ACS
pointings included less than half of the early-type Virgo members with
$B$ magnitude $B_T{\,<\,}16$ mag (though a complete sample of the
$\approx\,$20 brightest), and therefore the mapping of the early-types
was necessarily incomplete.  For complete areal coverage, one must
resort to wide-field imaging from the ground.  The only previous major
ground-based SBF survey was by \citet{tonry97,tonry01}, which reported
distances of variable precision, acquired in the $I$ band under
conditions of highly variable quality, for $\approx\,$300 galaxies out to
about 40~Mpc, including 31 in the Virgo cluster.  There have been
major advances in both CCD cameras and in the efficiency of
observatory operations since the observations were conducted for that
seminal survey more than 20~years ago; these advances greatly increase
the potential of the SBF method and make it well worth revisiting
today.  In fact, \citet{tonry01} hypothesized a future survey that
would repeat the SBF measurements for all the same galaxies in only
1/4 of the integration time and yield distances with 40\% better
precision because it would be conducted with a median seeing of
$\lta\,$0\farcs6, instead of $\approx\,$0\farcs9.


The Next Generation Virgo Cluster Survey \citep[NGVS;][]{ferrarese12} is a Large
Program with the 1-deg$^2$ MegaCam imager at the MegaPrime focus of the Canada-France Hawaii
Telescope (CFHT).  In 117 pointings (not including background fields), 
it covers a contiguous 104~deg$^2$ of the Virgo cluster, out to the virial radii
for both the Virgo~A and B subclusters, in the $u^*,g,i,z$ bandpasses.  
It supersedes all previous optical studies of Virgo and leverages a large amount of
spectroscopic follow~up and auxiliary data at other wavelengths to address important
questions about the galaxy luminosity function, scaling relations, stellar populations,
dynamical interactions, globular clusters, galactic nuclei,
and the growth of the cluster itself.
The NGVS project motivations, strategy, and observational program are
discussed in detail by \citet{ferrarese12}.

The NGVS observations were designed to deliver stacked images in the
$i$~band with seeing of 0\farcs6 or better to enable high-quality
measurement of the $i$-band SBF magnitude \mibar\ for the greatest
number of galaxies.  The wide-baseline photometry, from \ustar\ to
$z$, also enables accurate characterization of the galaxy
stellar populations, required for calibrating the absolute SBF
magnitude \Mibar.  The goal of the SBF component of the NGVS project
is to use the resulting distance moduli $(\mibar{-}\Mibar)$ to produce
the most detailed possible 3-D map of the cluster.  In the present
work, we present the first set of SBF measurements for 89 galaxies
brighter than $B_T{\,\approx\,}13$~mag. The following section briefly
describes the NGVS imaging and data reductions.
Section~\ref{sec_measure} details our SBF measurement procedures,
while in Section~\ref{sec_distances} we derive multiple calibrations
based on various photometric colors and apply the calibrations to
determine the galaxy distances. The distances are tabulated and
discussed in Section~\ref{sec_results}, where we also compare our
results with previous measurements and stellar population model
predictions.  Section~\ref{sec_summa} provides a summary.


\section{NGVS Data, Galaxy Sample, and Distance Zero Point}
\label{sec_data}
 
This work is based entirely on CFHT/MegaCam imaging data from the
NGVS.  Full details on the NGVS survey observations and image
processing are presented in \citet{ferrarese12}; here we provide only
a brief summary of the relevant details.  The NGVS exploits the
capabilities of CFHT/MegaCam to reach 5-$\sigma$ limiting magnitudes
for point source detection of 26.3, 26.6, 25.8, and 24.8~mag in the
stacked \ustar, $g$, $i$, and $z$ images, respectively, across the
entire Virgo cluster.  This is well beyond the turnover for the GCLF
\citep[e.g.,][]{durrell14}, even when the GCs are superimposed on a
bright galaxy background. The images are therefore well-suited for SBF
analysis, as most of the potentially contaminating sources can be
identified and removed.

Individual exposures in the NGVS survey were processed with
{Elixir-LSB}, a variant of the Elixir processing pipeline
\citep{elixir} specifically designed for the NGVS observing strategy,
that accurately removes background variations using sky frames
constructed from adjacent fields observed as part of a multi-pointing
``step-dither'' sequence.  The processed exposures for a given
pointing were then stacked using a variant of the MegaPipe
\citep{megapipe} pipeline.  In addition to removing scattered light,
this procedure effectively removes the sky fringing in the final
$i$-band images, and nearly eliminates it in $z$.  This is important
because residual fringing can contaminate the image power spectrum and
cause major problems the SBF analysis in ground-based data
\citep[e.g.,][]{tonry97}.  As a result of the excellent detrending and
sky subtraction with Elixir-LSB and the optimized stacking by
MegaPipe, the respective surface brightness limits (2-$\sigma$) in
\ustar, $g$, $i$, and $z$ are 29.3, 29.0, 27.4, and
26.0~mag~arcsec$^{-2}$.
As discussed in \citet{ferrarese12}, the image processing was somewhat
different for the central 4~deg$^2$ ``pilot region'', which was
observed before the final observing procedure was devised.  However,
experiments comparing our final results for the same fields using the
different processings showed negligible differences.

For this first paper of the NGVS-SBF project, we have selected the
complete sample of galaxies brighter than $B_T\approx13$ in the Virgo
Cluster Catalog (VCC) of \citet{binggeli85}, and falling within the
104~deg$^{2}$ NGVS footprint.  This initial sample of bright galaxies
has allowed us to optimize our SBF measurement procedures for
NGVS data and, as detailed in Sec.~\ref{sec_distances}, establish the
behavior of the $i$-band SBF magnitude over a broad range of
integrated galaxy colors.  The SBF analysis for a larger sample of
fainter galaxies is in progress and will be presented in a future
work.
Table~\ref{tab_all} lists the sample galaxies included in the present work.
For each galaxy, we give the VCC number from \citet[][]{binggeli85};
celestial coordinates (J2000) in degrees;
total $B_T$ magnitude from the VCC;
heliocentric velocity from NED;
ACSVCS SBF distance modulus \mMacs\ as tabulated by B09, when available;
morphological \Ttype\ from Hyperleda \citep[][]{makarov14};
and the alternative NGC and Messier names.
%

The availability of the \mMacs\ for $\approx\,$40\% of the sample
galaxies makes it possible to set the distance zero point for our NGVS
SBF measurements.  The mean distance modulus for the 85~galaxies in
the Virgo cluster proper (excluding the background \Wprime\ group and
the foreground NGC\,4697) from B09 is $31.092\pm0.013$, or
$16.5\pm0.1$~Mpc. This is the same mean distance as used in all
previous NGVS papers. The zero~point for the ACSVCS distances comes
from assuming the \citet{tonry01} mean distance for 31 Virgo galaxies,
revised by $-0.06$~mag \citep{blake02} based on comparison to the
final set of Key Project Cepheid distances from \citet{freedman01}.
With improvements in the precision of the Cepheid zero~point
\citep{freedman10}, the systematic uncertainty in this mean distance
is approximately 0.1~mag, or $\approx\,$0.8~Mpc, consistent with the
agreement with the predicted zero~point from the SPoT stellar
population models \citep{raimondo05,raimondo09}; see the discussion by
\citet[]{blake10b}.

\section{SBF measurements}
\label{sec_measure}

For measuring the SBF amplitudes in the sample galaxies, we adopt the same basic
procedures already developed in previous works and described in detail elsewhere
\citep{bva01,blake09,blake10b,cantiello05,cantiello07b,cantiello07a,cantiello11a,cantiello13,mei05iv,mei05v}, including both \hst\ and ground-based studies.
In broad terms, the SBF distance measurement entails: $(a)$~modeling
and subtracting of the 2-D galaxy surface brightness distribution and
the large-scale model residuals to obtain a clean residual image;
$(b)$~masking of contaminating sources (stars, background galaxies,
and especially GCs in the galaxy itself) down to a known threshold;
$(c)$~creating an accurate point spread function (PSF) template for
the image; $(d)$~determining the amplitude of the power spectrum in
Fourier space on the scale of PSF, which causes correlation of the
fluctuations in adjacent pixels; $(d)$~estimating the ``residual
variance'' from contaminating sources remaining in the masked image
and subtracting this variance from the power spectrum amplitude to
obtain the corrected SBF magnitude \mbar; $(e)$~adopting a value for
$\Mbar$ from either an empirical or theoretical SBF calibration
(generally based on galaxy color) to obtain the distance modulus,
$\overline{m}{\,-\,}\overline{M}$.

For this NGVS SBF analysis, we adopted the sky background estimates
and galaxy isophotal models described by \citet[][]{ferrarese12}.  The
large-scale residuals, still present in the frame after subtracting
the galaxy model, were removed using the background map obtained with
SExtractor \citep{bertin96} adopting a mesh size $\approx 10$ times the
FWHM \citep{tal90,cantiello05}.  We refer to image with the sky,
galaxy model, and large-scale residuals all subtracted as the {\it
  residual} frame.  The photometry of the external sources (stars,
background galaxies, and GCs) was done by running SExtractor on the
residual frame after masking saturated stars, very extended galaxies,
and other large features that could be problematic for SExtractor.  We
used an input weight image that included the galaxy photometric noise
as well as additional variance caused by the SBF
\citep[for details, see][]{jordan04,cantiello05} to avoid detecting
the fluctuations as objects themselves. The SExtractor MAG\_AUTO
measurements include roughly 90\% of the light from faint sources; the
aperture corrections for these magnitudes were obtained from a number
of isolated sources in each frame using a curve-of-growth analysis out
to large radii \citep[][]{cantiello09,cantiello11b}. The cutouts of
the NGVS images used for the analysis are $\approx5\arcmin$ wide each
side; on such scales the PSF variation is small enough that the
typical rms scatter is $\sim\,$0.01 mag for the aperture
corrections\footnote{For galaxies close to MegaPrime image tile edges,
  where the PSF or the image quality might be rapidly changing on
  spatially small scales, we limited our aperture corrections, PSF
  selection, and SBF analysis to the regions with the best image
  quality.}.

\begin{figure*}
  \vspace{-0.3cm}
  \hspace*{-0.1cm}\includegraphics[scale=.204]{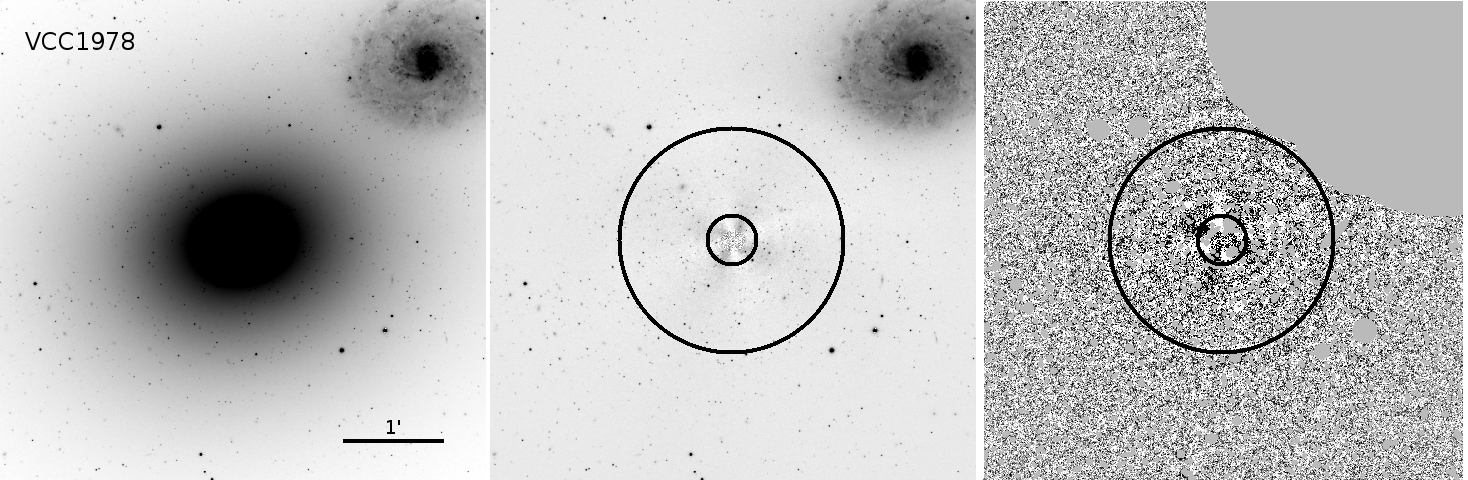}
  \includegraphics[scale=.394]{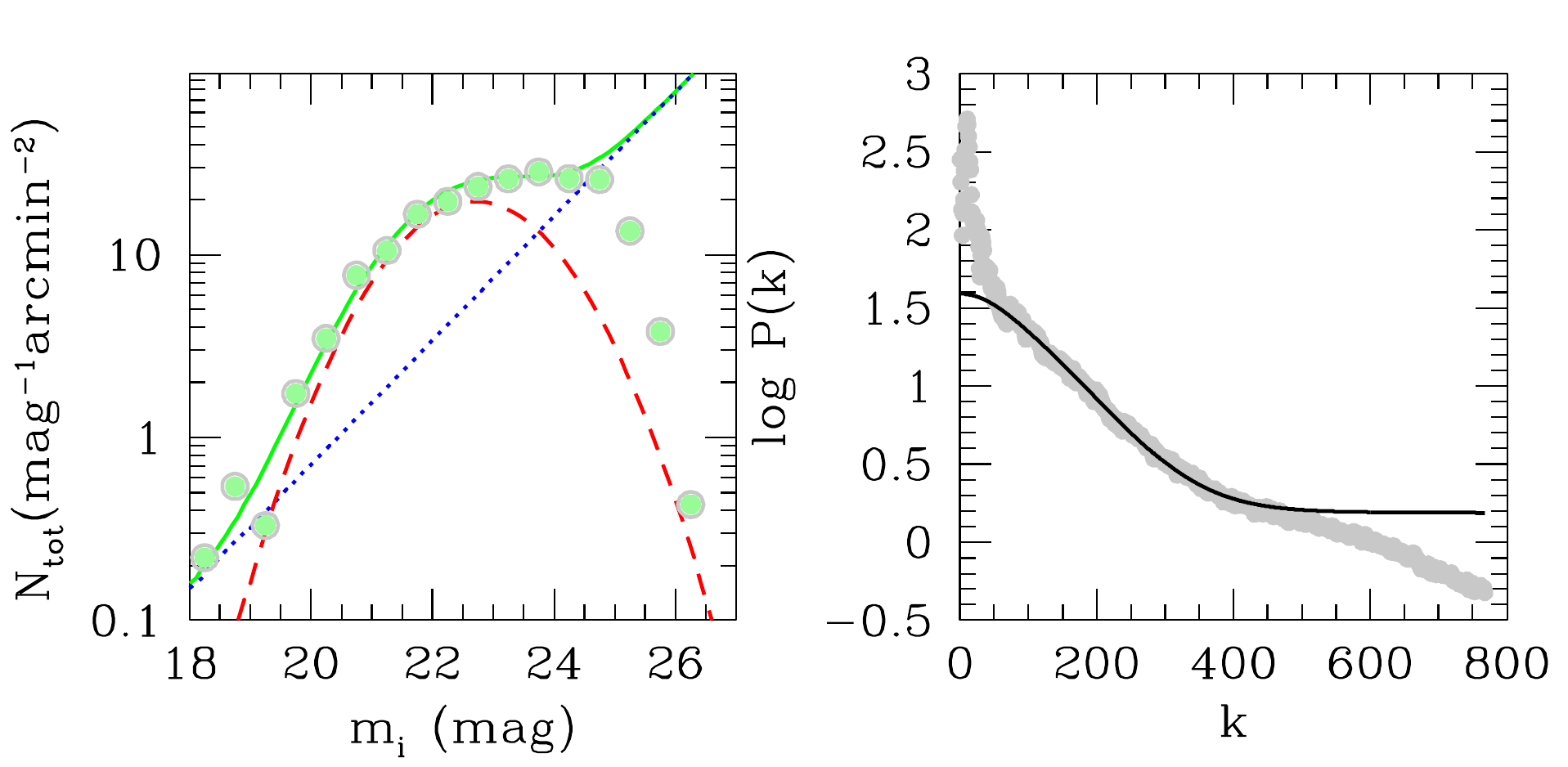}\hspace*{+0.1cm}
  \\
  \hspace*{-0.1cm}\includegraphics[scale=.204]{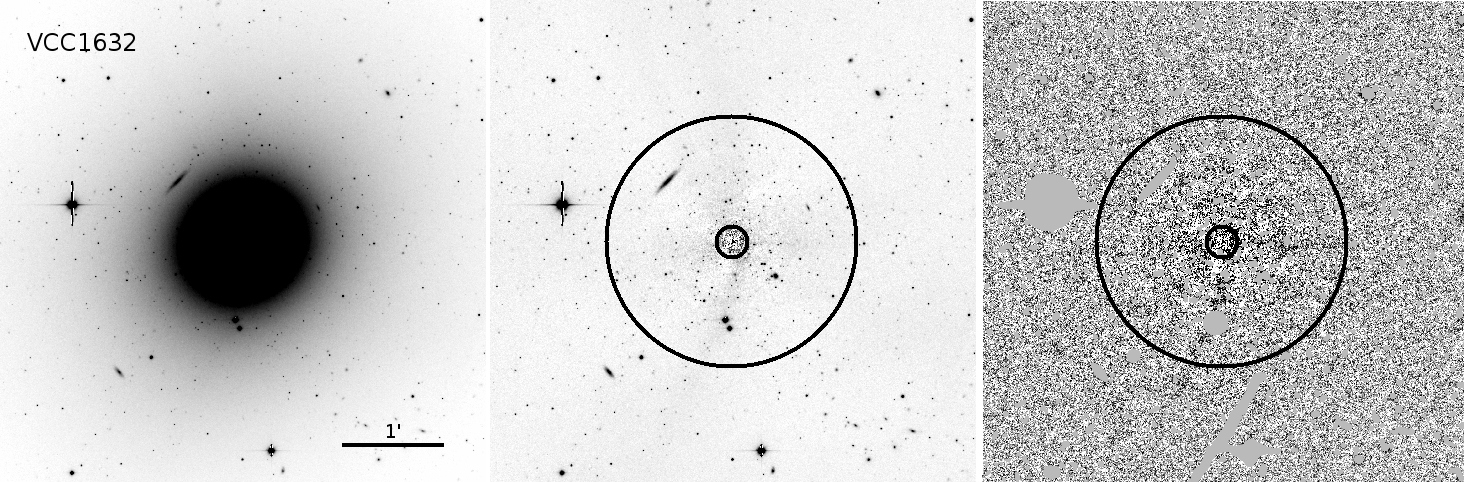}
  \includegraphics[scale=.394]{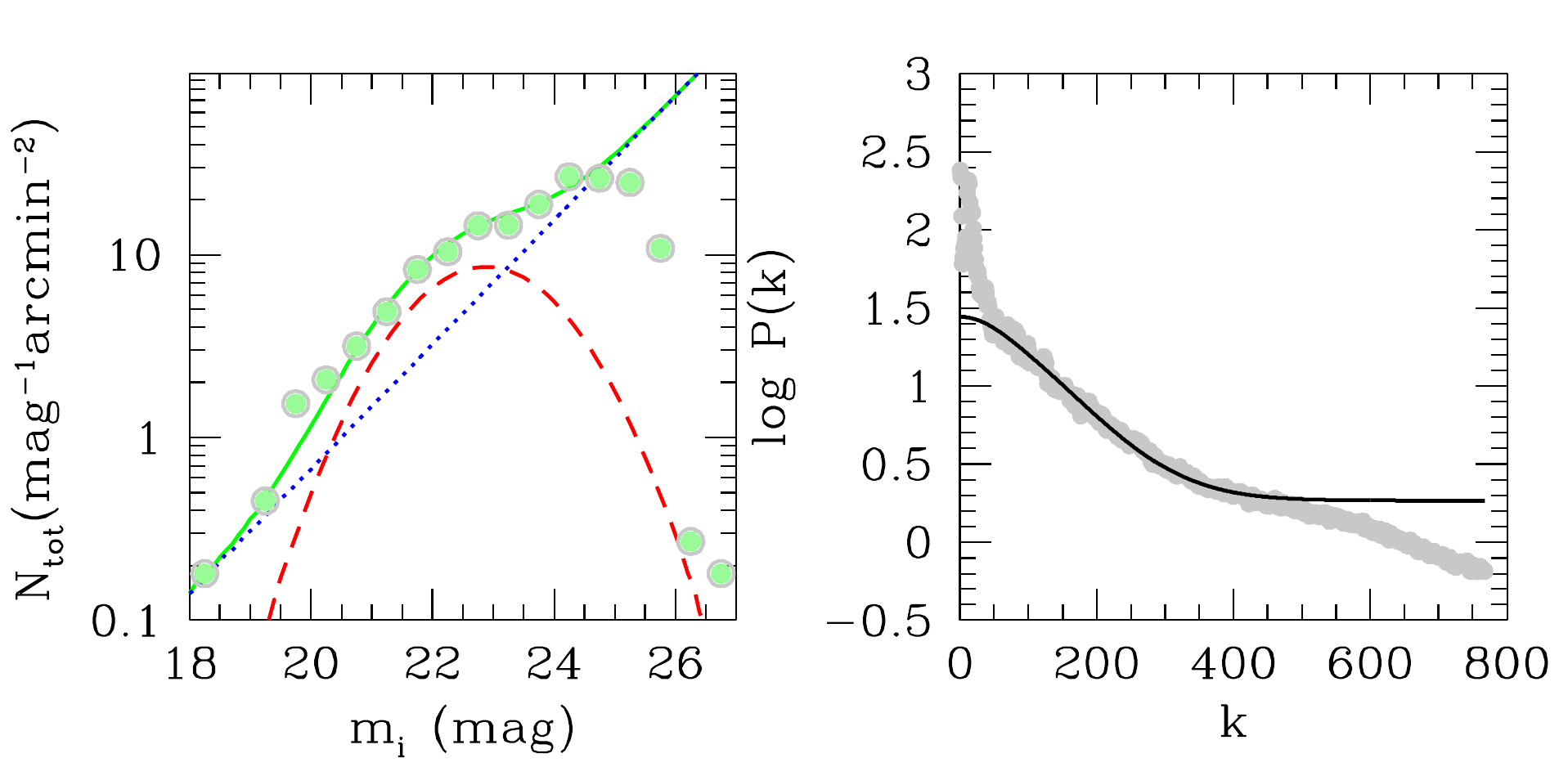}\hspace*{+0.1cm}
  \\
  \hspace*{-0.1cm}\includegraphics[scale=.204]{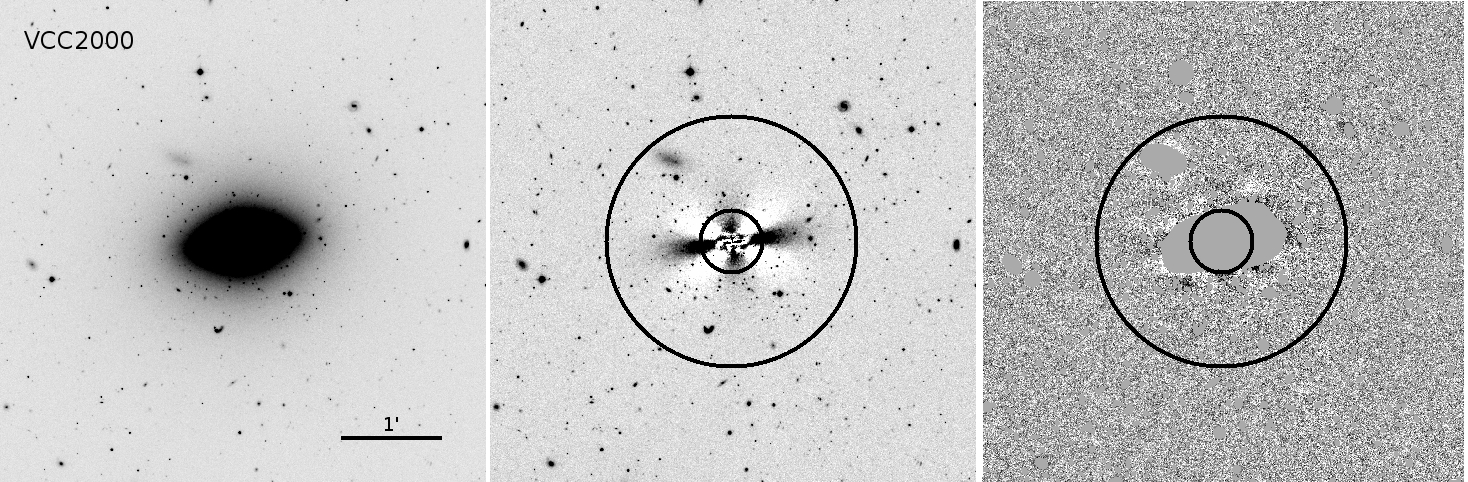}
  \includegraphics[scale=.394]{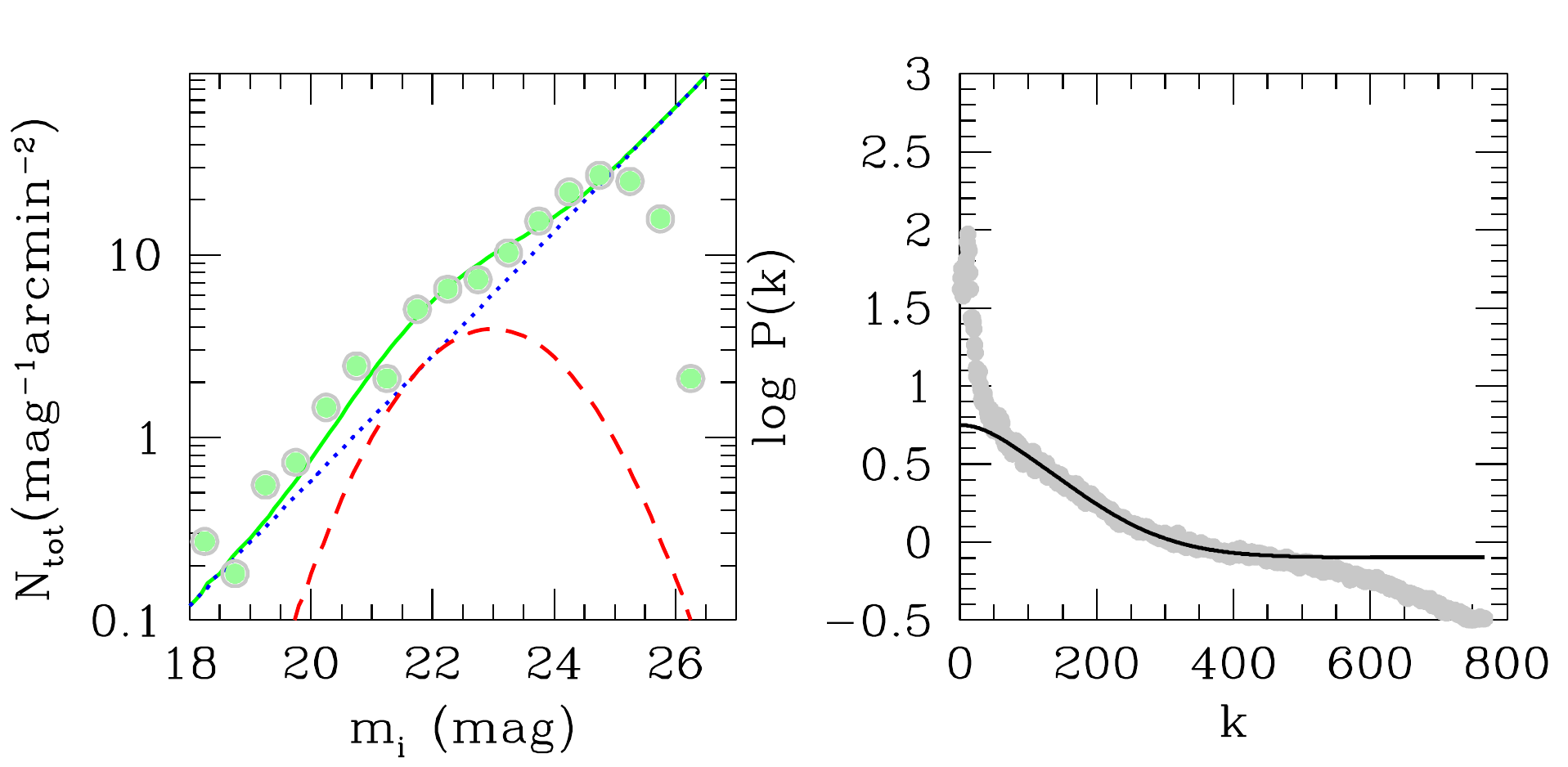}\hspace*{+0.1cm}
  \\
  \hspace*{-0.1cm}\includegraphics[scale=.204]{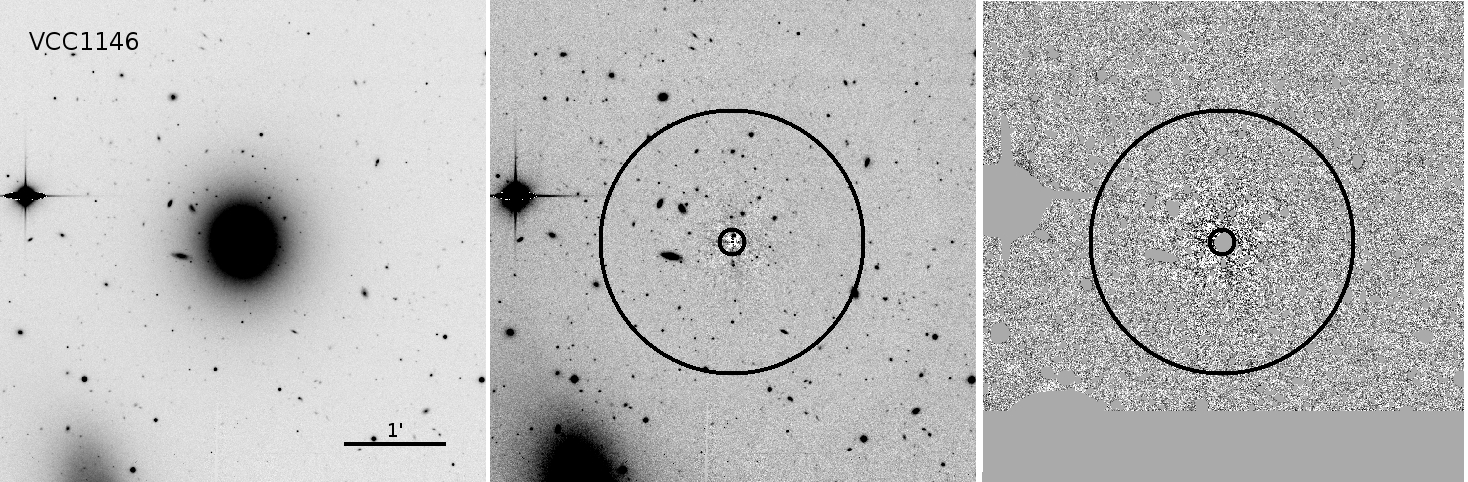}
  \includegraphics[scale=.394]{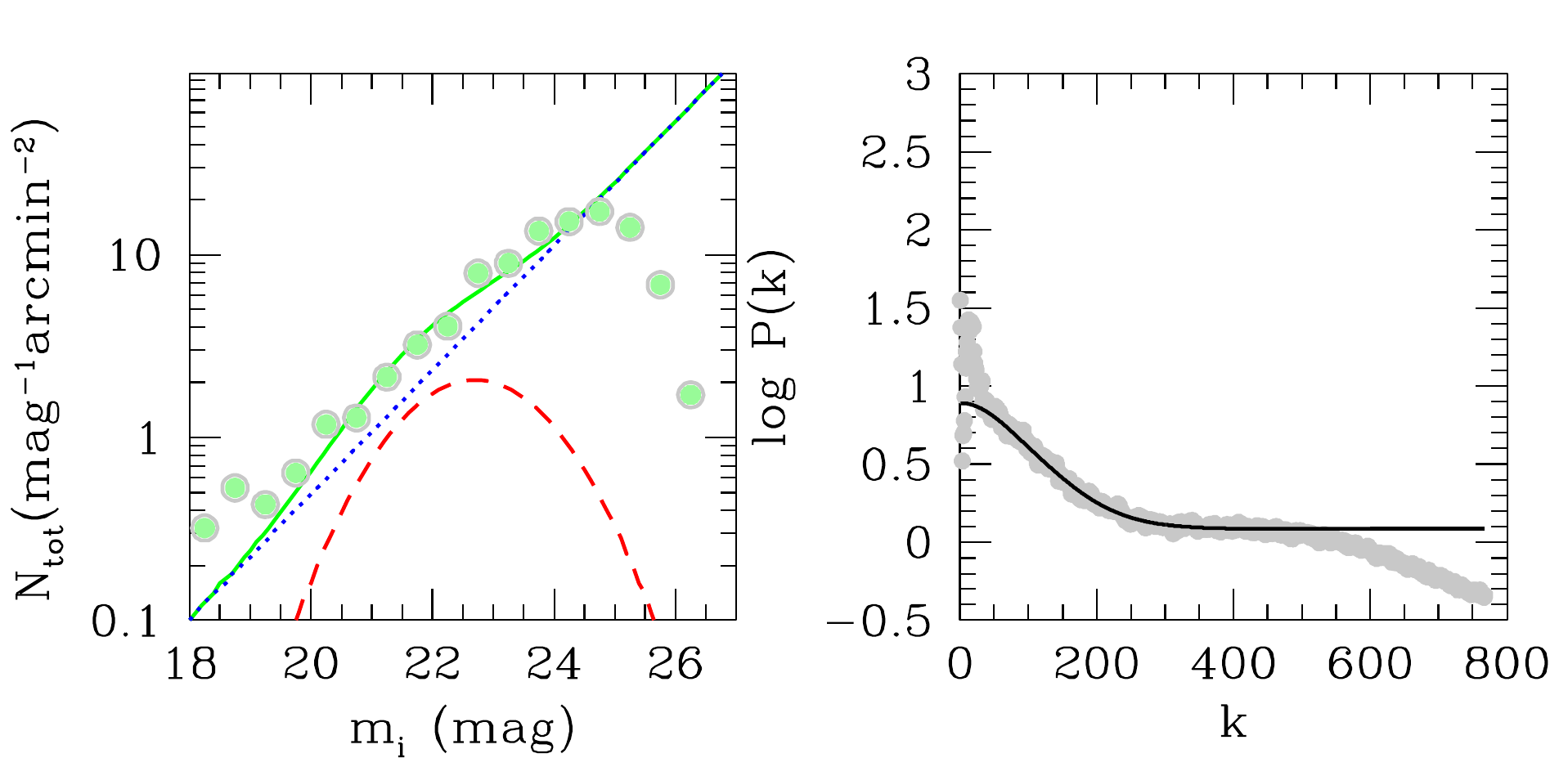}\hspace*{+0.1cm}
  \\
 \caption{SBF analysis images and plots for a selection of galaxies,
   as labeled. Starting from the left: $i$-band image, residual, and
   residual masked image (first to third panel). The black annuli in
   the second and third panels show the inner and outer radii of the
   region adopted for SBF measurements. Fourth panel: Fitted
   luminosity function of external sources. Filled green circles mark
   observational data; the solid green curve is the best fit to the
   data; the two components of the total luminosity function, i.e. the
   background galaxies and the GCLF, are shown with a blue dotted and
   a red dashed curve, respectively. Fifth panel: Azimuthal average of
   the residual image power spectrum (gray dots) and the fit obtained
   according to the procedure described in text (solid black curve).
\label{pofit}}
\vspace{0.4cm}
\end{figure*}

Once the catalog of sources had been derived, the next step was to fit
the luminosity function of the sources, necessary in order to estimate
the amount of contamination from unmasked faint sources in the
residual frame.  We fitted a source magnitude distribution to a model
including a combination of the GCLF and a power-law background galaxy
luminosity function (examples are shown in Figure~\ref{pofit},
discussed below).  As noted above, bright stars and the most extended
sources had been masked and were therefore not present in the
catalog. At these high Galactic latitudes, the surface density of
fainter stars is small compared the GCs in the vicinity of bright
galaxies. The best fit to the combined GCLF and galaxy luminosity
function was used to derive background fluctuation correction term,
$P_r$, as described in previous works
\citep[e.g.,][]{tal90,mei05v,cantiello05}.

To determine the SBF amplitudes, we measured the azimuthal average of
the power spectrum $P(k)$ within circular annuli of the masked
residual frame, and modeled $P(k)$ as the power spectrum of a template
PSF convolved with the mask image, $E(k)$, plus another term
representing the power spectrum of noise unconvolved with the PSF.
For the PSF term, we used from two to twelve individual isolated
bright point sources near the target galaxy in each residual frame, as
well as model PSFs constructed with DAOPhot
\citep{stetson87,stetson90}.  Each PSF was normalized and used
separately to estimate the total fluctuation amplitude $P_0$ via a
robust minimization method \citep{press92} as the multiplicative
factor in the power spectrum representation $P(k)=P_0 \times E(k) +
P_1$, where $P_1$ is the unconvolved ``white noise'' term.  We
averaged the values of $P_0$ determined from all the available PSF
templates, rejecting any PSFs that gave poor fits to the power
spectrum.  Finally, the SBF amplitude $P_f$ was found by subtracting
the background contamination term from the power spectrum amplitude
$P_f=P_0-P_r$.

Figure \ref{pofit} illustrates the basic steps involved in the SBF
measurement for four example galaxies representing the quartiles of
the $B_T$ interval for the sample studied in this work.  In each row
of the figure, the panels show: the target galaxy image; residual
frame at the same intensity stretch with sources unmasked; residual
frame with a tighter stretch after masking detected sources; fitted
background luminosity function model; power spectrum of the residual
frame compared to the scaled PSF power spectrum.  The upturn in the
power spectrum at low wavenumber $k$ occurs because of remaining
large-scale features in the residual frames; the downturn at high $k$
occurs because of correlation by the sinc-like interpolation kernel
during image stacking \citep[see][]{mei05iv,cantiello05}.  These high
and low $k$ ranges are excluded from the power spectrum fits. To
define the best interval for the $k$ wavenumbers, we examined the
residuals of the observed power spectrum with respect to the model and
rejected $k$-numbers where the residuals start deviating
systematically from zero \citep[e.g., Figure 7 in][]{cantiello13}. The
exact $k$ interval adopted for the fits depends on the size of the
image used for the SBF analysis.

\begin{figure*}
%
\begin{center}
  \vspace{-0.5cm}
  \includegraphics[scale=.3]{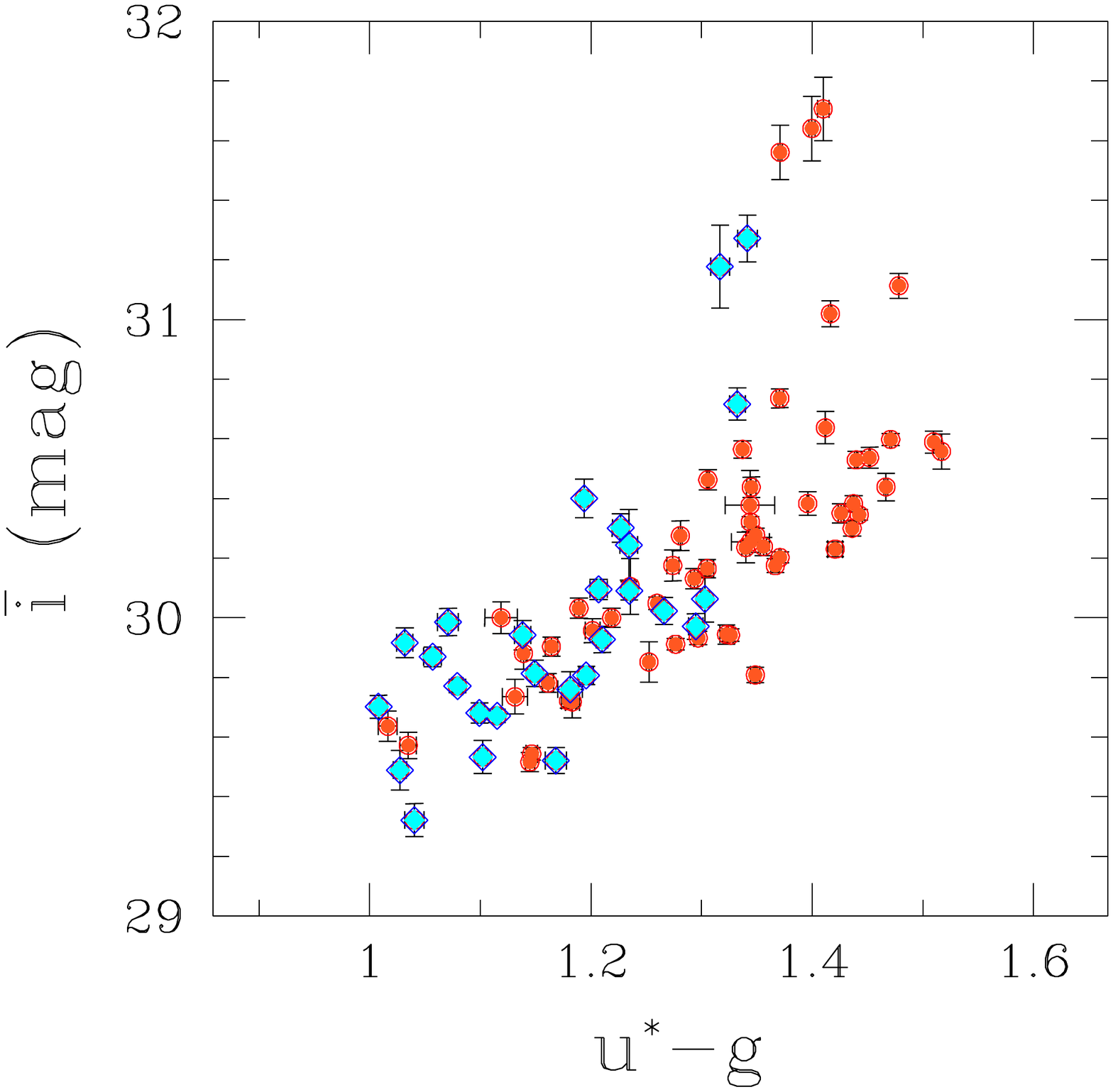} \vspace{-2.3cm}
  \includegraphics[scale=.3]{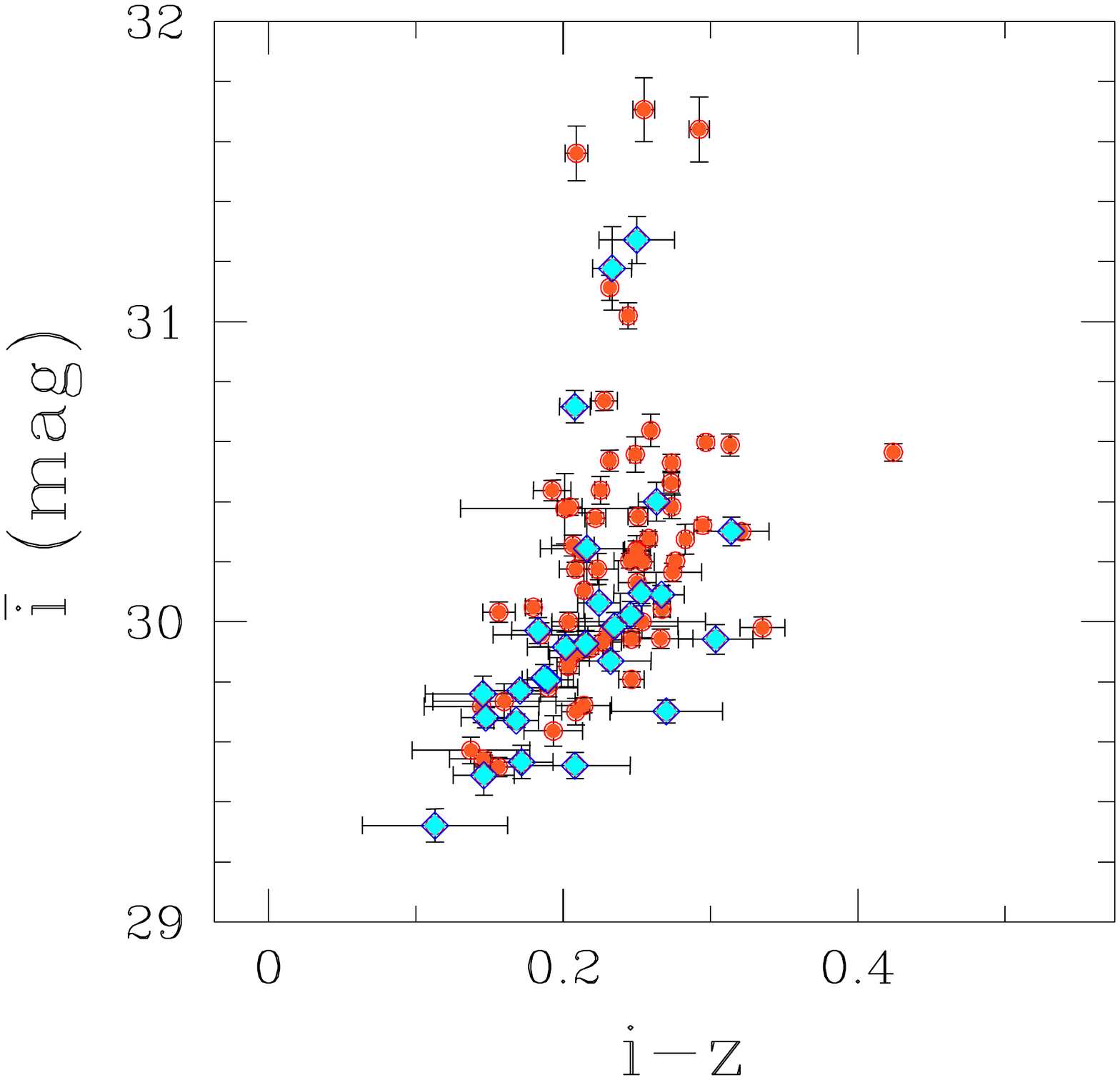}
  \\ \includegraphics[scale=.3]{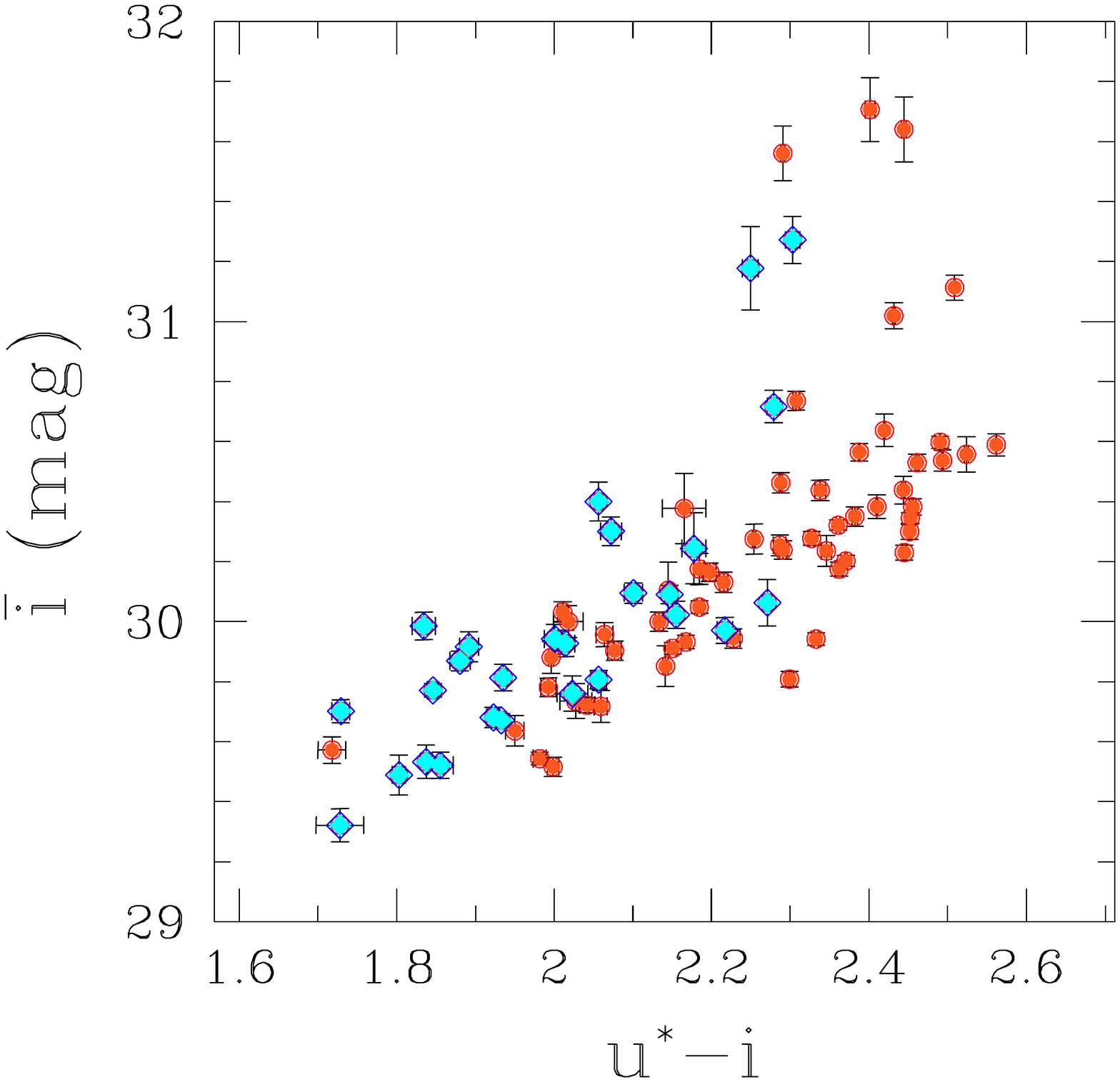} \vspace{-2.3cm}
  \includegraphics[scale=.3]{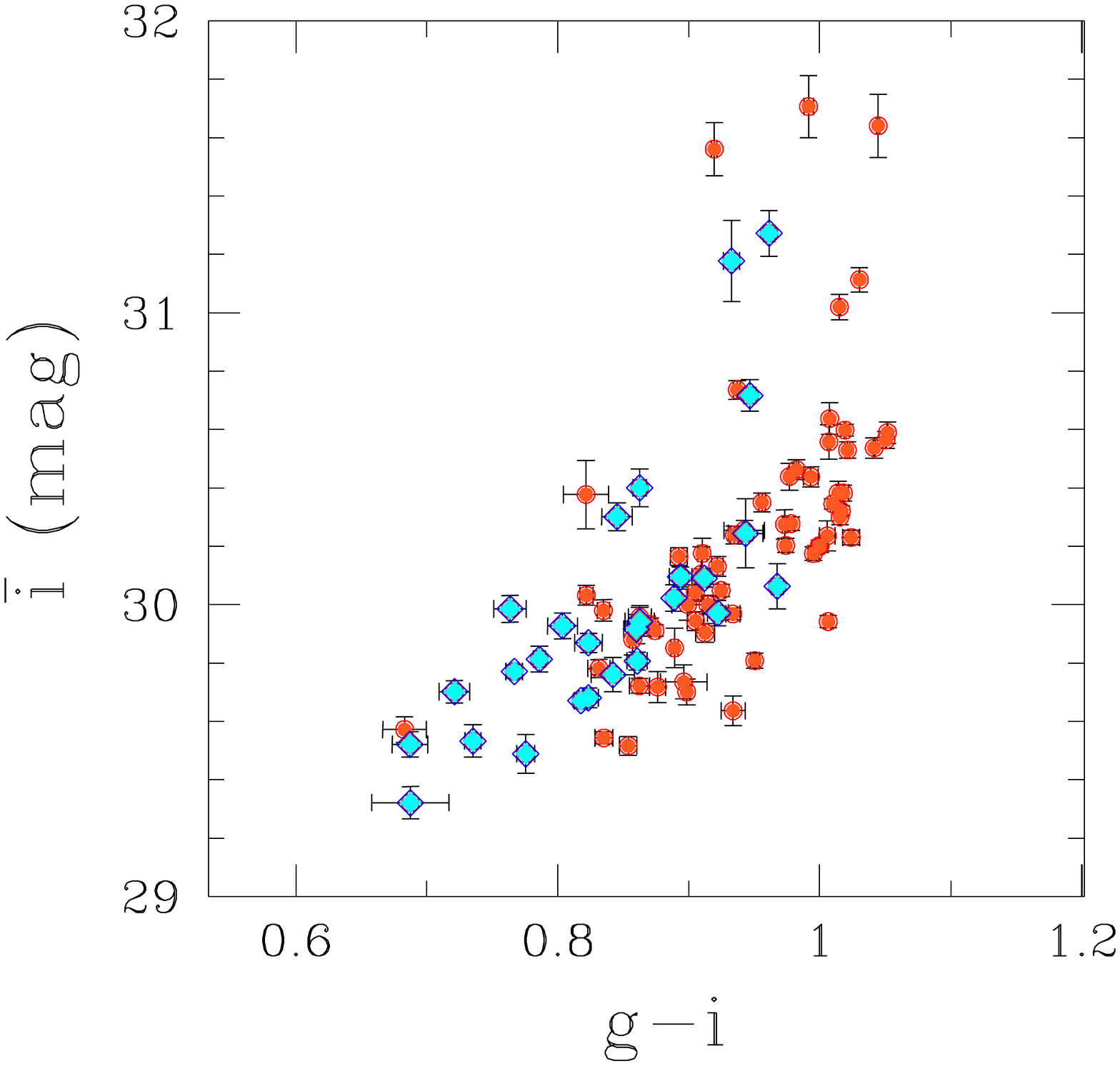}
  \\ \includegraphics[scale=.3]{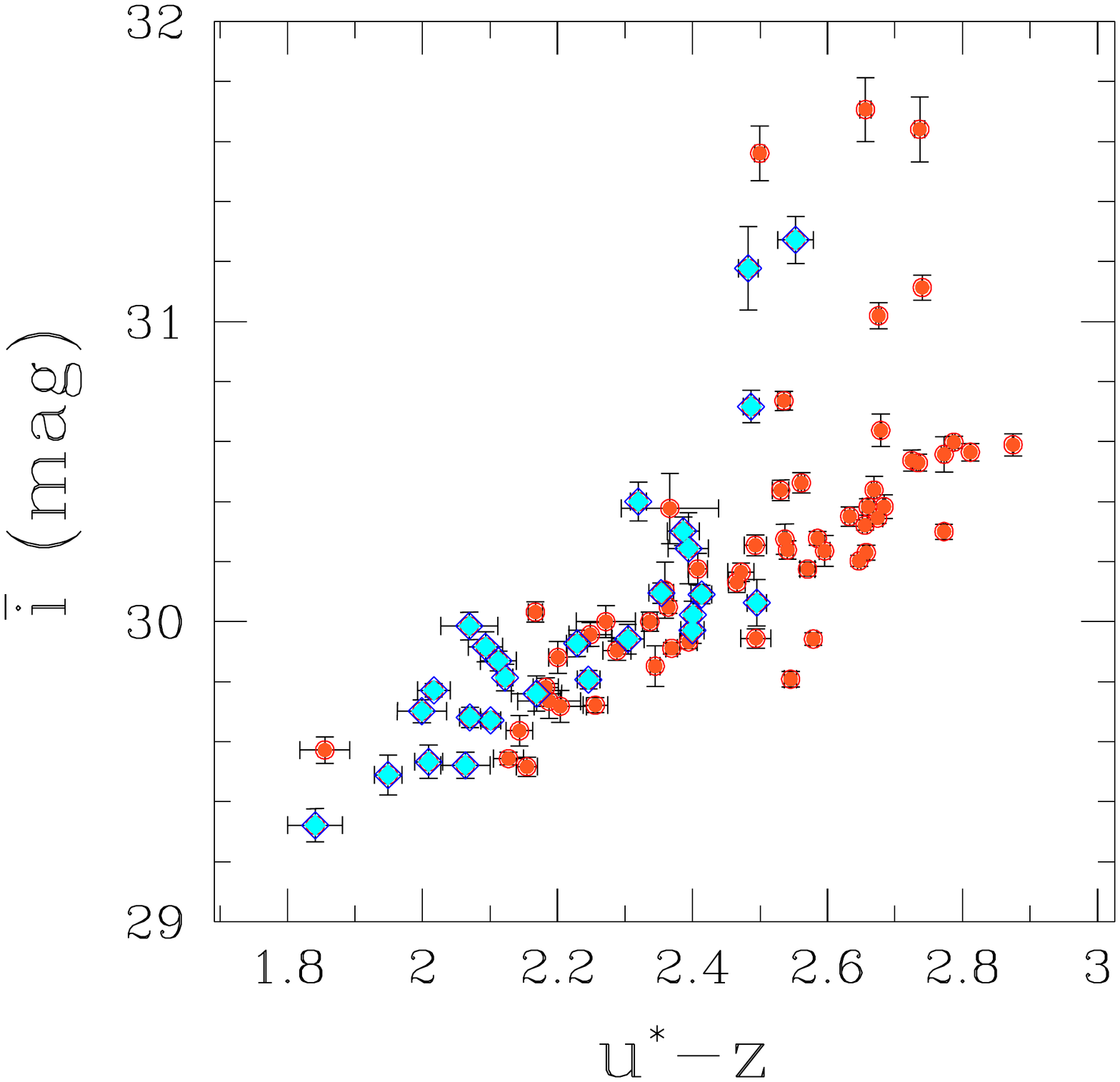} \vspace{-1.8cm}
  \includegraphics[scale=.3]{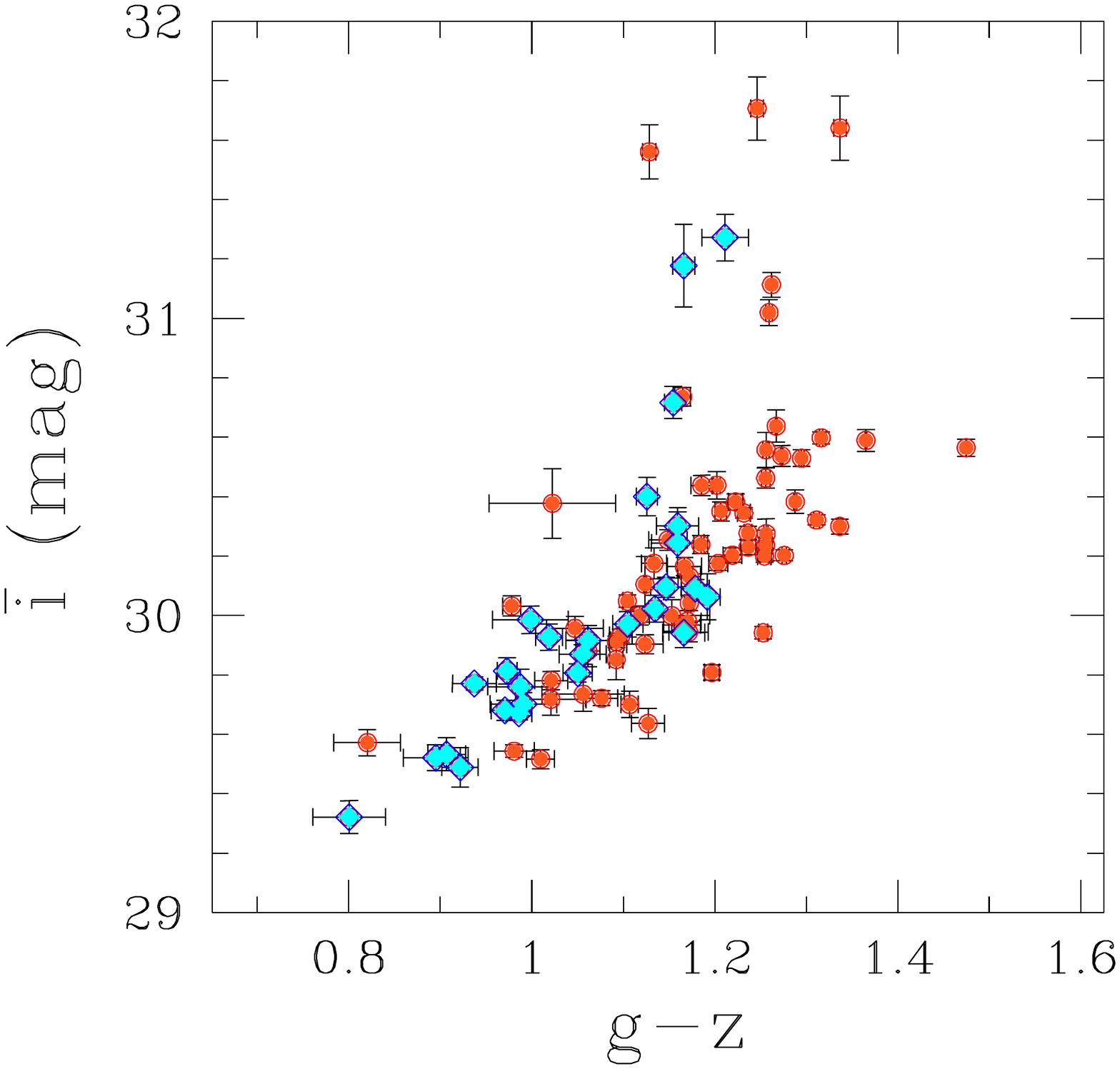}
\end{center}
\caption{Measured SBF amplitudes versus color. Red circles mark
  early-type galaxies, blue diamonds show late type galaxies.
\label{ibarcol}}
\end{figure*}

The second and third panels of each row in Figure~\ref{pofit} indicate
the annuli used for the SBF measurements in the example galaxies.  We
also measure the integrated colors of the galaxies in the same annuli
as used for the SBF measurements.  Table~\ref{tab_sbf} reports the
area and the median radius of the annulus adopted for each galaxy in
our sample, along with all possible unique color measurements and the
SBF magnitude \mibar\ derived within the annulus\footnote{In order to
  derive the distance moduli $(m{-}M)$, the colors and magnitudes
  reported in Table \ref{tab_sbf}, uncorrected for Galactic
  extinctions, were corrected using the values from \citet{sfd98}.}.
As explained in the following section, the table also includes our
preferred distance modulus for each galaxy and a comment about the
likely subgroup membership inferred from the galaxy distance modulus.
Figure \ref{ibarcol} plots the apparent SBF magnitudes for the full
sample of galaxies reported in Table~\ref{tab_all} versus various
integrated colors. In the panels of this figure, the early-type
galaxies with morphological $\Ttype{\,<\,}0$ and late-type galaxies
with $\Ttype{\,\geq\,}0$ are represented with red circles and blue
diamonds, respectively.

\begin{figure*}
\includegraphics[scale=0.45]{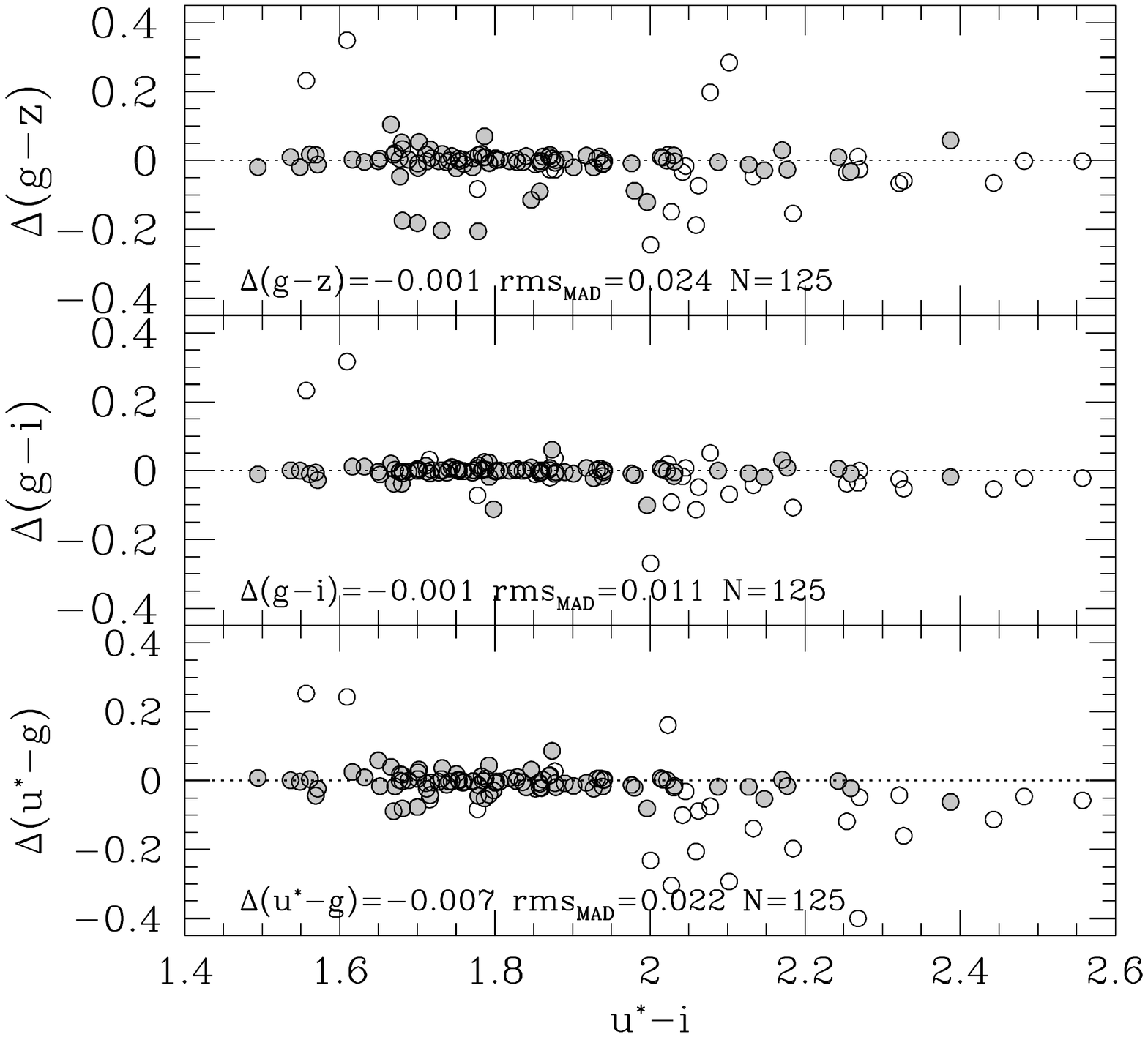}\hspace{-0.8cm}
\includegraphics[scale=0.45]{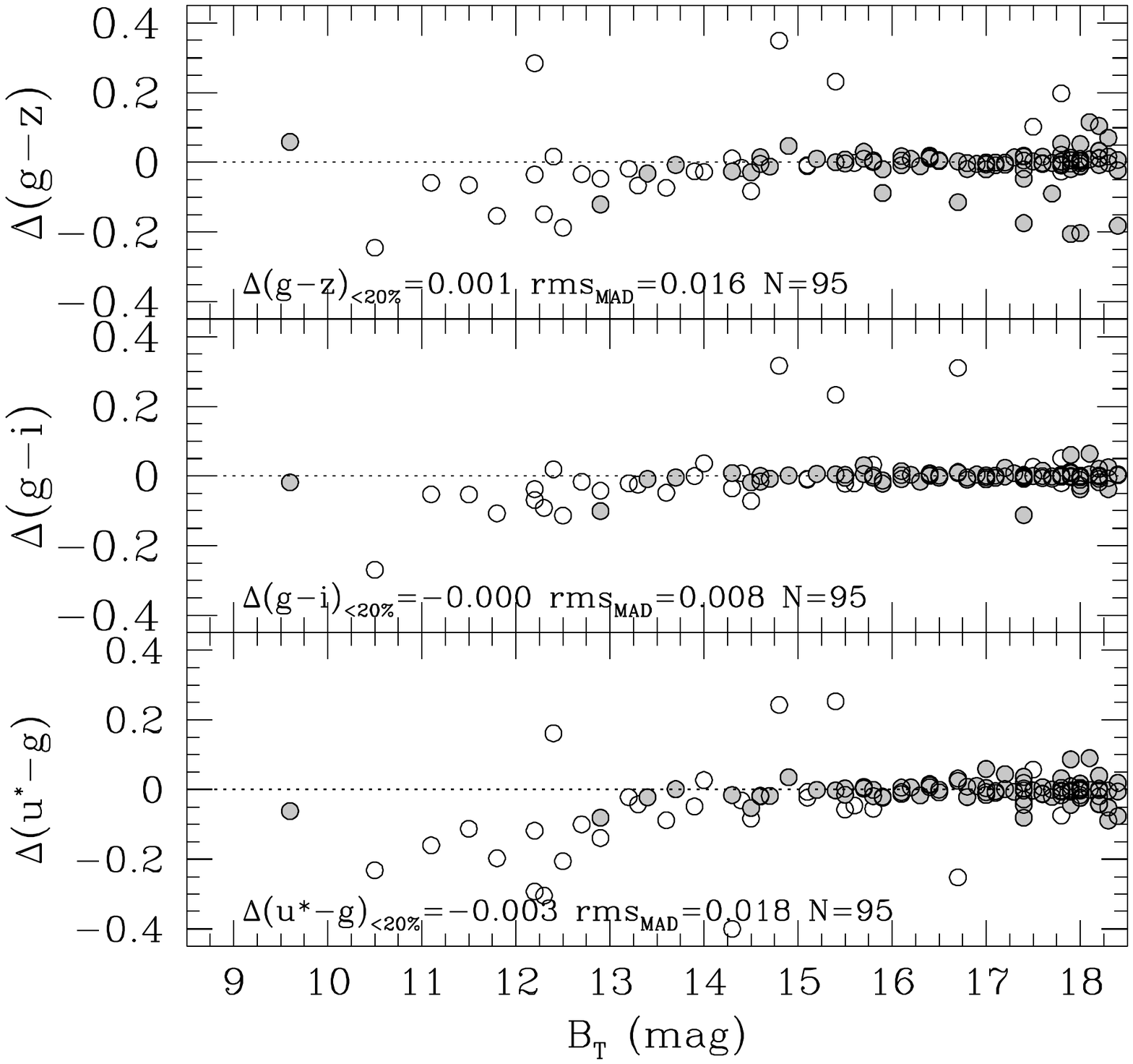}\vspace{-2.4cm}
\caption{Left panels: color differences between the present work and
  \citet[][see text]{roediger17}. Filled dots refer to objects with
  SBF measurement radius within 20\% the effective radius adopted by
  \citet{roediger17}. Right panels: as left, but versus total
  magnitude $B_T$. The median difference
  ($\Delta(X{-}Y)=(X{-}Y)_\mathrm{this~work}-(X{-}Y)_\mathrm{Roediger+17}\,$),
  rms, and the number of objects used for the full sample of common
  sources is reported in left panels. The same quantities for the
  selected objects with SBF and $\Reff$ radii matching within 20\%,
  are given in right panels.
\label{joel}}
\end{figure*}

\section{Analysis}
\label{sec_distances}

To derive distances from the SBF measurements, an accurate calibration
of \Mbari\ is required. In this section we analyze various options for
the calibration and choose the optimum approach.  The resulting
distances for the 89 galaxies in our sample are presented in
Section~\ref{sec_results}.

\subsection{Calibrating SBF: distances and colors}

The measured SBF magnitudes show a clear dependence on galaxy color
that is recognizable even before applying any cluster-depth
correction.
Clearly, some data points in the plots are outliers with respect to
the visual mean relations: this is in part due to the combination of
the intrinsic depth of Virgo and the intrinsic scatter of the SBF
method itself, but also to the presence of a sequence of objects
offset by $\approx+0.7$ mag with respect to the most populated sequence
on the \mbari-versus-color diagrams, associated with the Virgo
\Wprime\ group, and to a fraction of objects which are background
galaxies, possible members of the W and M subgroups
\citep{sandage85,binggeli87,binggeli93,kim14}.
To derive absolute \Mbari\ magnitudes, we adopt the ACSVCS distances
listed in Table~\ref{tab_all} for the 36 galaxies in common with our
sample.

To verify our colors measurements, we compared our results with the
ones from \citet[][]{roediger17}, obtained from the same NGVS data
used here. To increase the number of galaxies for the comparison, we
extended the present dataset to galaxies fainter than the limiting
$B_T$ adopted in this paper, and for which SBF analysis is in
progress.
Figure \ref{joel} shows the color difference \citep[this work minus
  the results from][]{roediger17} versus \ui\ (left panels) and versus
total magnitude $B_T$ (right panels). Although based on the same
dataset, the statistical consistency of the measurements is not
trivial, because the two studies adopt different analysis procedures,
and the annular regions as well as the masking thresholds are
different. The galaxies with the largest mismatch are the ones where
the mean radius of the region inspected for SBF measurements differs
substantially from the galaxy effective radius, $\Reff$, adopted by
\citet{roediger17} to determine colors. The comparison limited to the
galaxies with $|(\Reff-R_{SBF})/R_{SBF}|\leq0.2$ (that is the ones
where the median radius of the region used for SBF and $\Reff$ differ
less than $\approx20\%$) shows remarkable agreement, with
$\Delta~color<0.01$ mag and root-mean-squared scatter
$\mathrm{rms}\lsim0.02$ mag\footnote{We also compared our colors with
  the measurements from \citet{chen10} based on SDSS data. The
  agreement is generally satisfactory for objects where the radius
  adopted for SBF analysis is similar to the annulus used for
  measuring colors.}.

\begin{figure*}
\includegraphics[scale=0.84, bb=0 200 585  600]{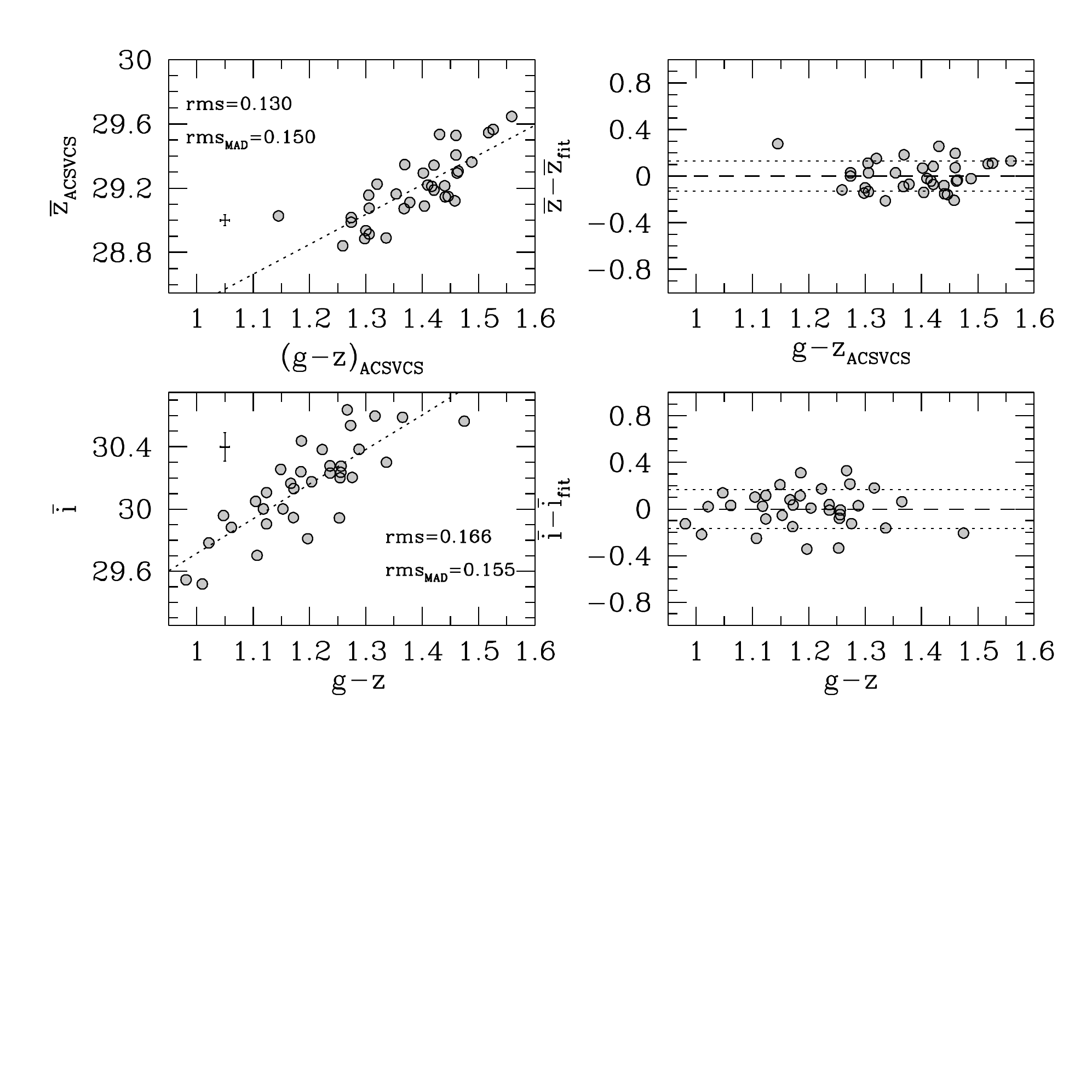}
\caption{Left panels: measured SBF amplitudes versus \gz\ from the
  ACSVCS ($z-$band, upper panel) and from the present work ($i-$band,
  lower panel). The linear least squares fit to the data is reported
  with a dotted line. The rms scatter, and \rmsmad\ with respect to
  the linear fit are reported in each panel. The median error bars are
  also reported in the panels. Right: residuals with respect to the
  linear fit for the ACSVCS (upper panel) and the NGVS (lower). The
  dashed and dotted lines represent the zero and $\pm$1-$\sigma$
  levels, respectively.
\label{compacal}}
\vspace{0.4cm}
\end{figure*}

\begin{figure*}
\vspace{-2cm}
\includegraphics[scale=0.84]{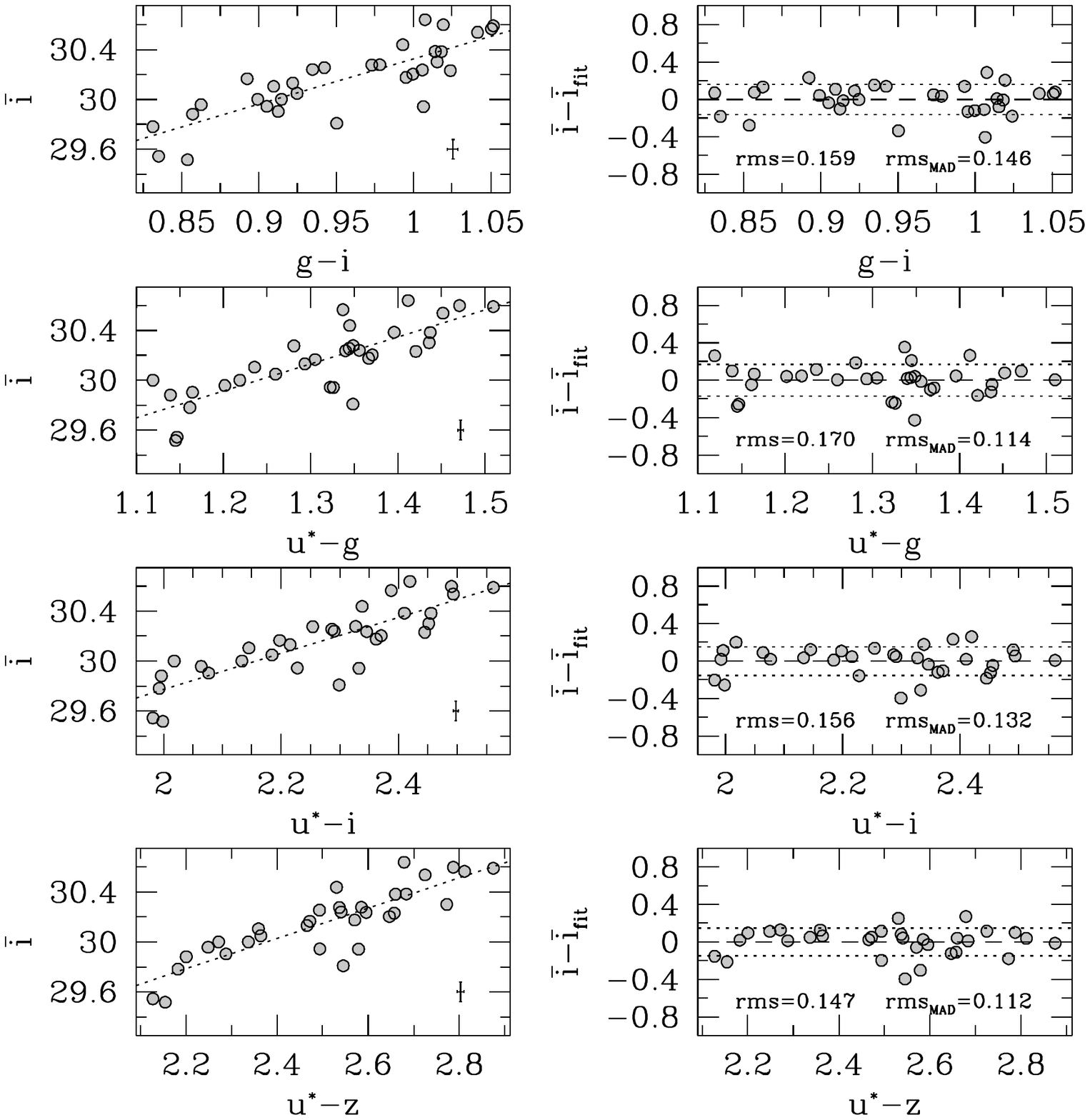}
\vspace{-5cm}
\caption{As in Figure \ref{compacal}, but for a selection of the other
  colors available from NGVS.
\label{compacal_ngvs}}
\vspace{0.5cm}
\end{figure*}

\subsection{ACSVCS $z$-band SBF versus NGVS $\overline{m}_i$}

A direct comparison of SBF magnitudes for the 36 galaxies in common
between this work and the ACSVCS is indicative of the extra
uncertainty of the present dataset due to the lower spatial resolution
of ground-based measures from CFHT/MegaCam data with respect to the
space-based HST/ACS data.

Figure \ref{compacal} shows the apparent SBF magnitude versus
\gz\ (left panels), and the residual with respect to a linear fit
(right plots) for both the ACSVCS (upper panels) and NGVS measurements
(lower panels)\footnote{The ACSVCS team used a more refined
  calibration scheme than the simple linear scheme used here. However,
  the ACSVCS sample covered a wider color range,
  $0.8\leq \gzacsvcs \leq$1.6 mag.  Within the narrower color interval
  of the present work, the proximity to a linear relation increases,
  as the bend of the $\overline{M}_z$ calibrations by
  \citet{mei07xiii} and B09 appears close to $\gzacsvcs\approx1.3$ mag.}.
The linear regressions, reported with dotted line in left panels of
the figure, are obtained after rejecting from the sample the two
galaxies in the \Wprime\ group (i.e., VCC\,731 and VCC\,1025). The rms
reported in the figure shows that the set of measurements from the
NGVS has $\approx0.04$ mag larger scatter than the ACSVCS. The scatter
estimate $\rmsmad{\,\equiv\,}1.48{\times}\mathrm{MAD}$, derived from
the median absolute deviation (MAD\footnote{The median absolute
  deviation, defined as $MAD=median|X_i-median(X)|$, is equal to the
  standard deviation for a Gaussian distribution, but is more robust
  than the standard deviation as it is less sensitive to outliers.}),
is similar to the rms.

Figure \ref{compacal_ngvs} shows the apparent magnitudes
\mbari\ versus colors other than the \gz. For the sake of comparison,
the two galaxies without currently available $u$-band magnitudes are
not included in the plots.  Together, Figures~\ref{compacal}
and~\ref{compacal_ngvs} show that, over the color interval of our
bright galaxy sample, a linear relation is generally a good
representation of the variation in SBF magnitude with color.  However,
even with the \wprime\ galaxies omitted, depth effects within Virgo
serve to increase the observed scatter in these relations (see Section
\S \ref{sec_results}).  We therefore move on to fitting the
calibrations in terms of absolute SBF magnitude.

\subsection{Single-color calibrations}

For our first set of calibrations, we derive linear fits to the
variation of the \Mbari\ values (derived from our \mbari\ measurements
and the ACSVCS distance moduli listed in Table~\ref{tab_all}) using a
single NGVS color index as the independent variable.  Since two of the
36 galaxies with ACSVCS distances lack \ustar~data, we therefore use
34 galaxies for these fits (including here the somewhat more distant
\Wprime\ members).
These single-color fits are parameterized as
\begin{equation}
  \overline{M}_i \;=\; \alpha \,+\, \beta\times[color - \mathit{refcolor}]\,.
  \end{equation}
The fitted $\alpha$ and $\beta$ parameters, as well as the adopted
``reference color'' (\textit{refcolor}), are given in
Table~\ref{tab_1col} for four different color indices.

\begin{figure*}
\vspace{-2.0cm}
\includegraphics[scale=0.84]{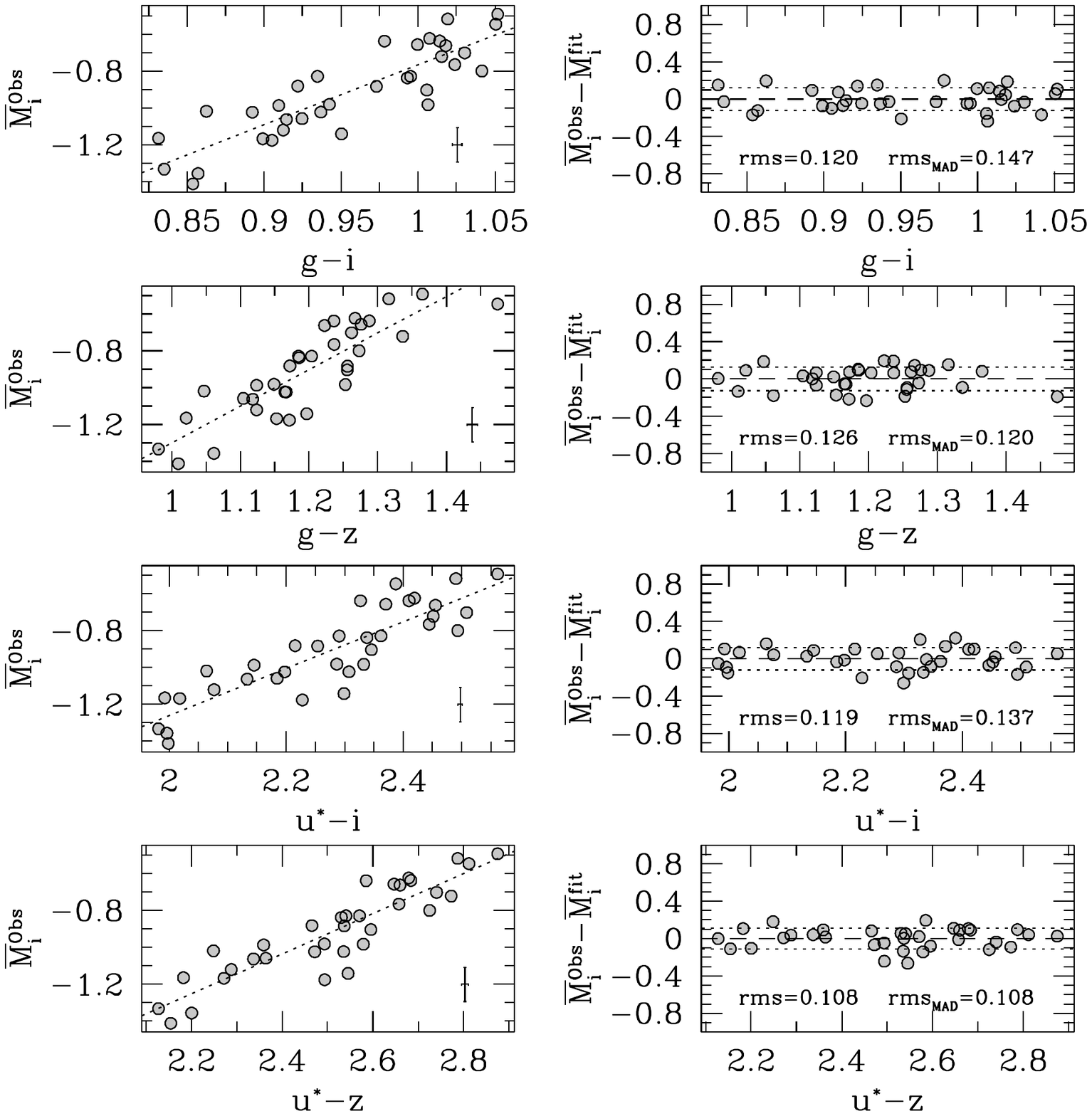}
\vspace{-5.0cm}
\caption{Left panels: Absolute SBF magnitudes, $\overline{M_i}^{Obs}$,
  versus color. The linear fit to data is shown with a dotted line;
  the median error bars include the uncertainty on the assumed
  distance modulus. Right panels: residuals with respect to the linear
  fit. The zero and $\pm$1-$\sigma$ lines are shown with dashed and
  dotted lines, respectively. The rms and \rmsmad\ for each are
  reported in right panels.
\label{colfit4}}
\vspace{0.5cm}
\end{figure*}

The four single-color linear fits are shown in Figure~\ref{colfit4}.
The left panels plot the absolute SBF magnitude versus 
measured color along with the linear fit to the data (dotted lines);
the right show the residuals with respect to the fits and
the measured scatter values.  Over the color range of the galaxies in
this sample, the fit of $\overline{M}_i$ versus
\uz\ (bottom panels in Figure \ref{colfit4}) exhibits the lowest
scatter, with $\mathrm{rms}\approx0.11$~mag.

\begin{figure*}
\includegraphics[scale=0.84]{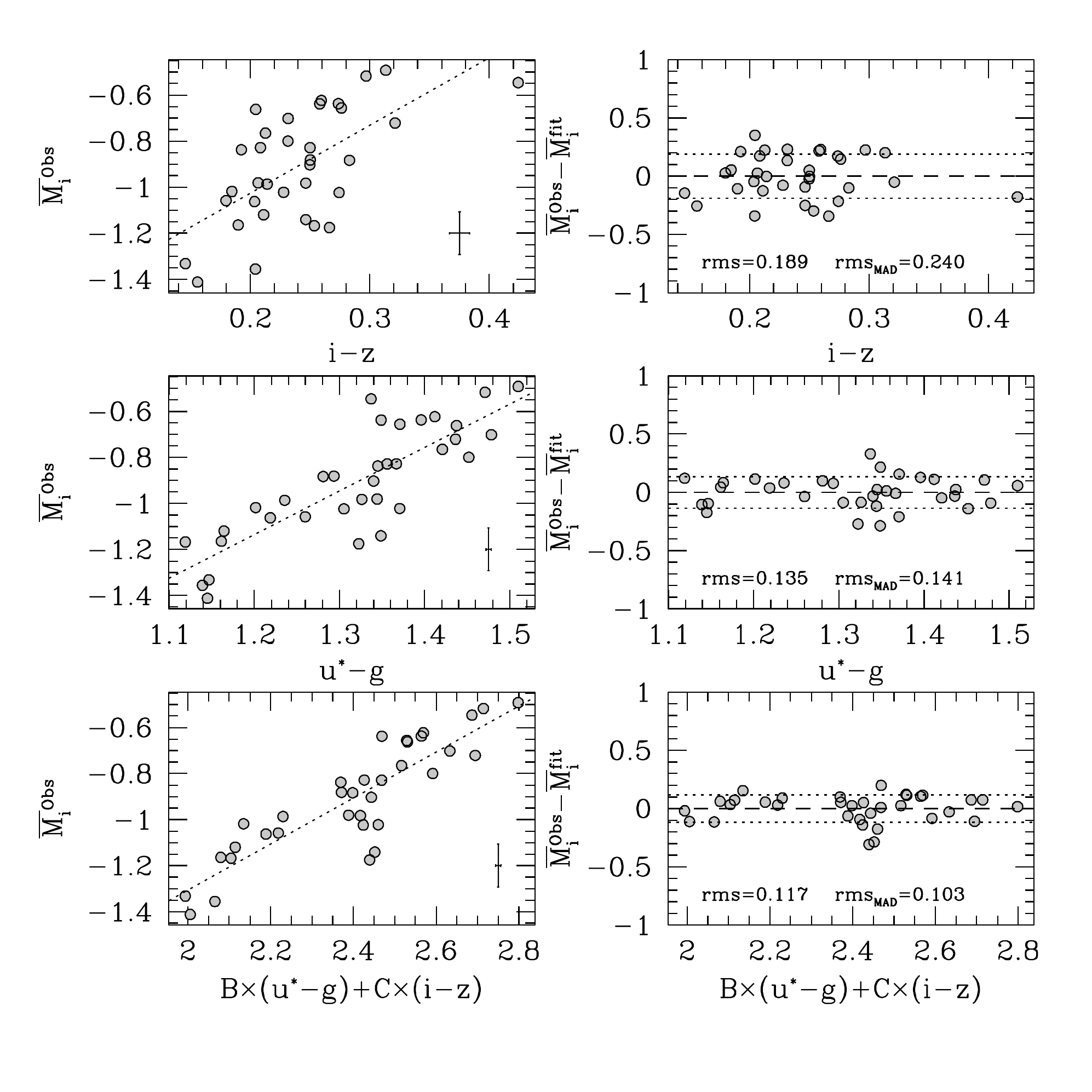}
\vspace{-1.0cm}
\caption{Upper and middle left panels: SBF versus color relations for
  \ug\ and \iz. The residuals with respect to the linear regression
  are shown in right panels, using the same symbols as in Figure
  \ref{colfit4}. The scatter of \ug\ and \iz\ color relations is
  sensibly larger than for the other colors, shown in Figure
  \ref{colfit4}. Lower panels: two-color fit (left panel) and
  residuals (right) obtained by combining the \ug\ and \iz. Note the
  lower scatter for the two-color with respect to the single color
  relations.
\label{2colfit_ug_iz}}
\vspace{0.5cm}
\end{figure*}

\begin{figure*}
\vspace{-2cm}
\includegraphics[scale=0.84]{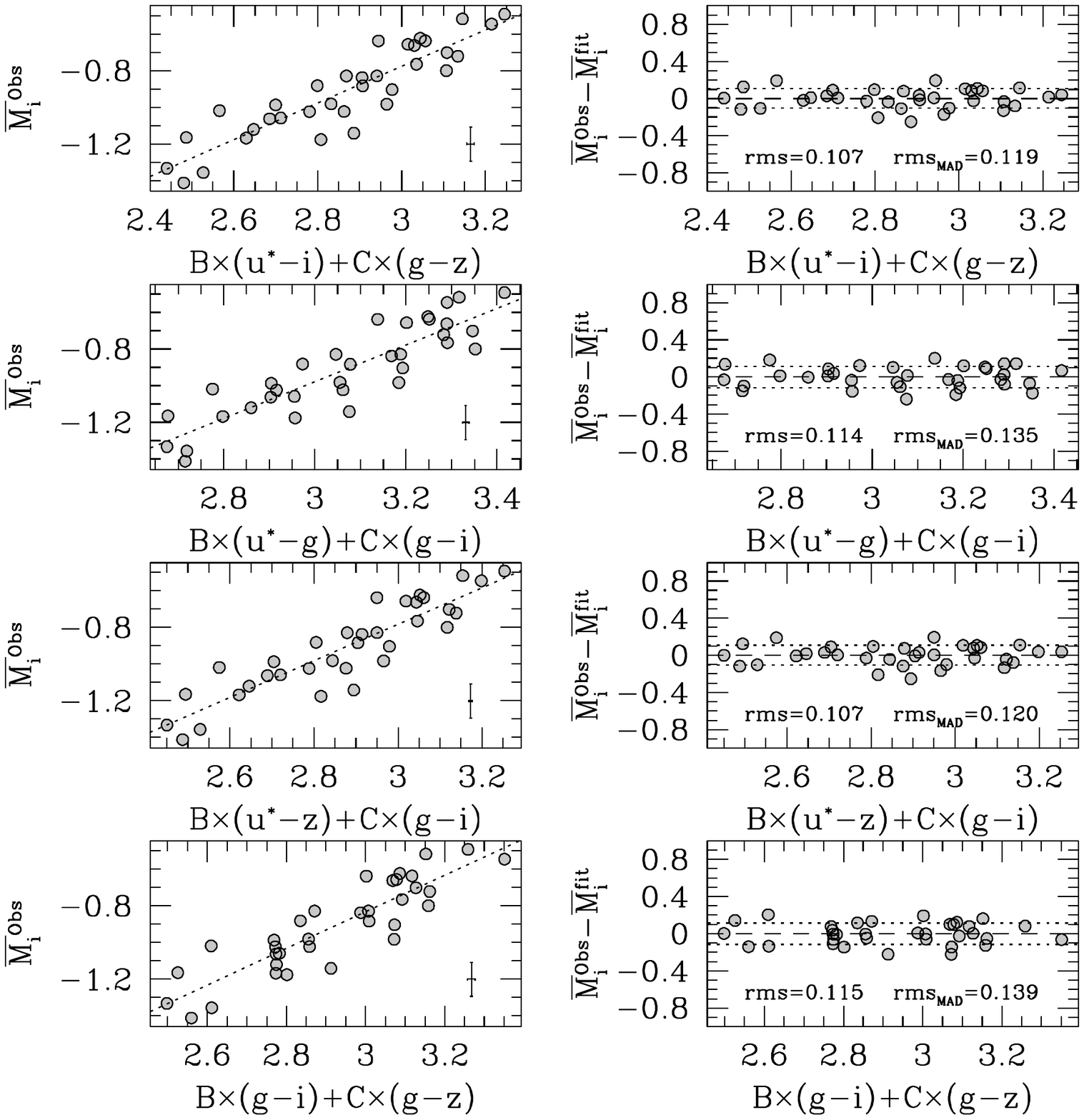}
\vspace{-5cm}
\caption{Two-color calibrations for various color
  combinations. Left/Right panels show the data used for the fits and
  the residual with respect to the latter.
\label{2colfit_all}}
\vspace{0.5cm}
\end{figure*}

\subsection{Dual-color calibrations}

SBF calibrations based on multiple color indices can potentially
provide improved characterization of the dependence of $\overline{M}_i$ on
stellar population properties. This would manifest as a lower scatter in the
empirically derived SBF calibrations.
For example, Figure \ref{2colfit_ug_iz} shows one case where the combination 
of two color indices, \ug\ and \iz, gives a lower calibration scatter than 
either of the colors individually
(single-color fits and residuals are shown in the upper and middle panels;
the lower panels show the dual-color relation).
However, because of their narrow wavelength baselines, neither 
\ug\ nor \iz\ are preferred choices for the single-color calibration.
Thus, this case is simply an illustration of the concept.

Using the available NGVS color data, we derived fits of $\overline{M}_i$ to
combinations of two different color indices in the following form:
\begin{equation}
  \overline{M}_i \;=\; A \,+\, B{\times}\mathit{color}_1 +\, C{\times}\mathit{color}_2 .
\end{equation}
The fitted values of $A$, $B$, and $C$ for four different choices
of $color_1$ and $color_2$ are given in Table \ref{tab_2col}.
Figure~\ref{2colfit_all} shows the four dual-color fits
and their residuals.

In general, by comparing the \Mbari\ calibrations based
on a single color with the calibrations based on a 
combination of two colors,  we find that even the widest
two-color combinations result in a 
calibration scatter for \Mbari\ comparable to the single-color \uz\
calibration. That is to say, for the present magnitude-limited sample,
we do not find strong evidence in favor of the two-color option
with respect to the more standard one-color $\overline{M}_i$ calibration,
as long as the calibrating color involves the \ustar~band.

Nevertheless, for the forthcoming larger  set of NGVS SBF measurements,
extending to fainter galaxies that tend to be bluer than the color range 
explored in the current sample, the two-color calibrations confer some
benefits over the traditional single-color option.
For instance, using a single color index for the calibration,
B09 showed that the scatter increased at the bluest colors; models 
suggest that this scatter may be reduced with a second color,
although this remains to be confirmed empirically.
Moreover, the simple linear approximation to SBF-color relation is
no longer valid at such blue colors \citep[][B09]{mei05v};
thus, the calibration will need to be revisited for the extended sample 
in any case.

\begin{figure}
  \begin{center}
    \includegraphics[scale=0.42]{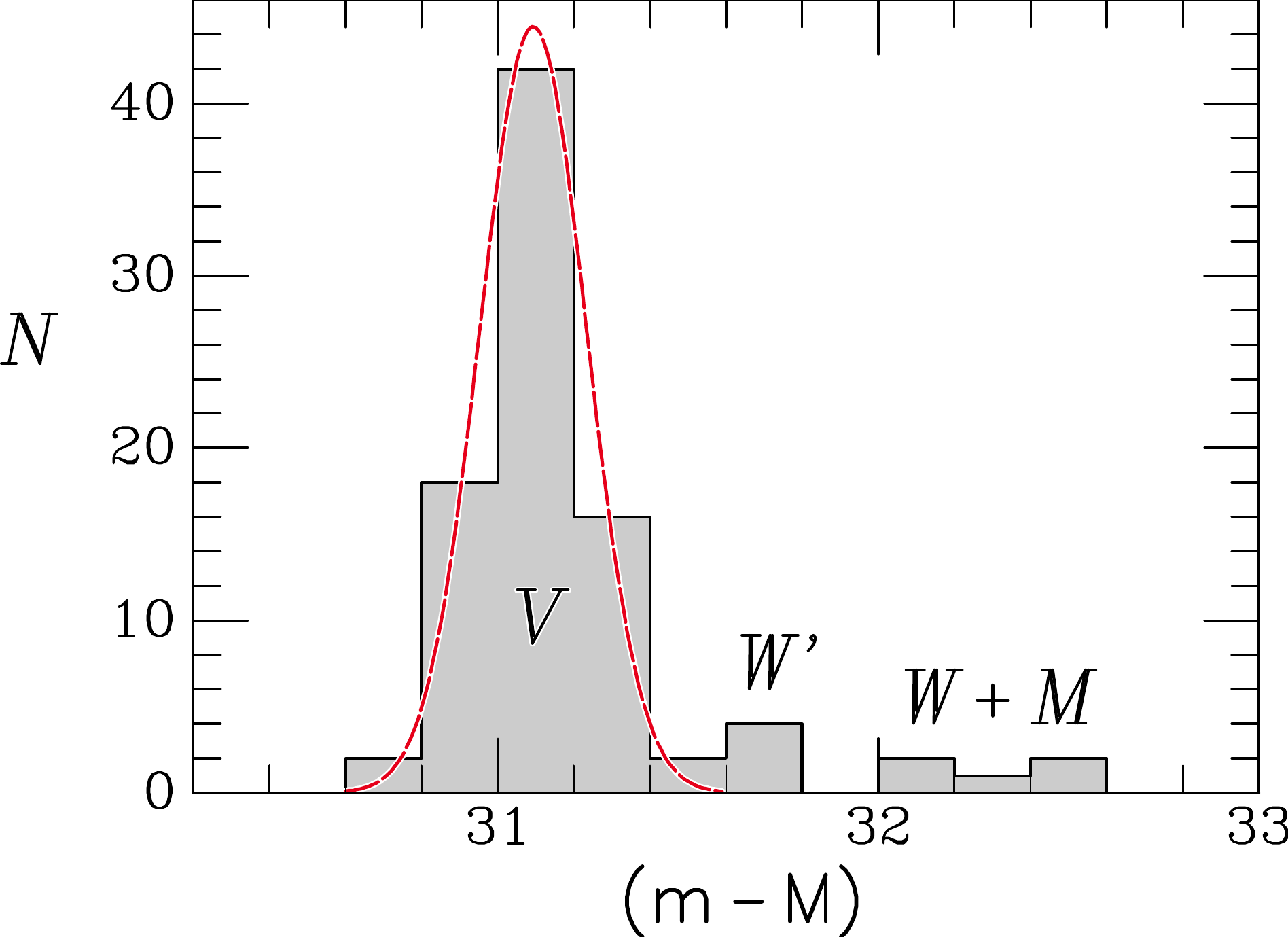}
\caption{Histogram of distance moduli presented in Table~\ref{tab_sbf}. The loci
  of the main Virgo cluster (V in the plot), 
  the \Wprime\ group, and the W and M clouds are indicated.
  For comparison, the red dashed curve shows  a Gaussian of
  mean = 31.09~mag and $\sigma=0.14$~mag, which provides a reasonable representation
  of the peak corresponding to the main Virgo cluster.
  Note that 0.14~mag is both the rms dispersion that we measure and the median
  estimated error; thus the width of the V peak is mainly due to measurement error.
\label{virgoers}}
\end{center}
\end{figure}

\begin{figure*}
\begin{center}
\includegraphics[scale=0.4]{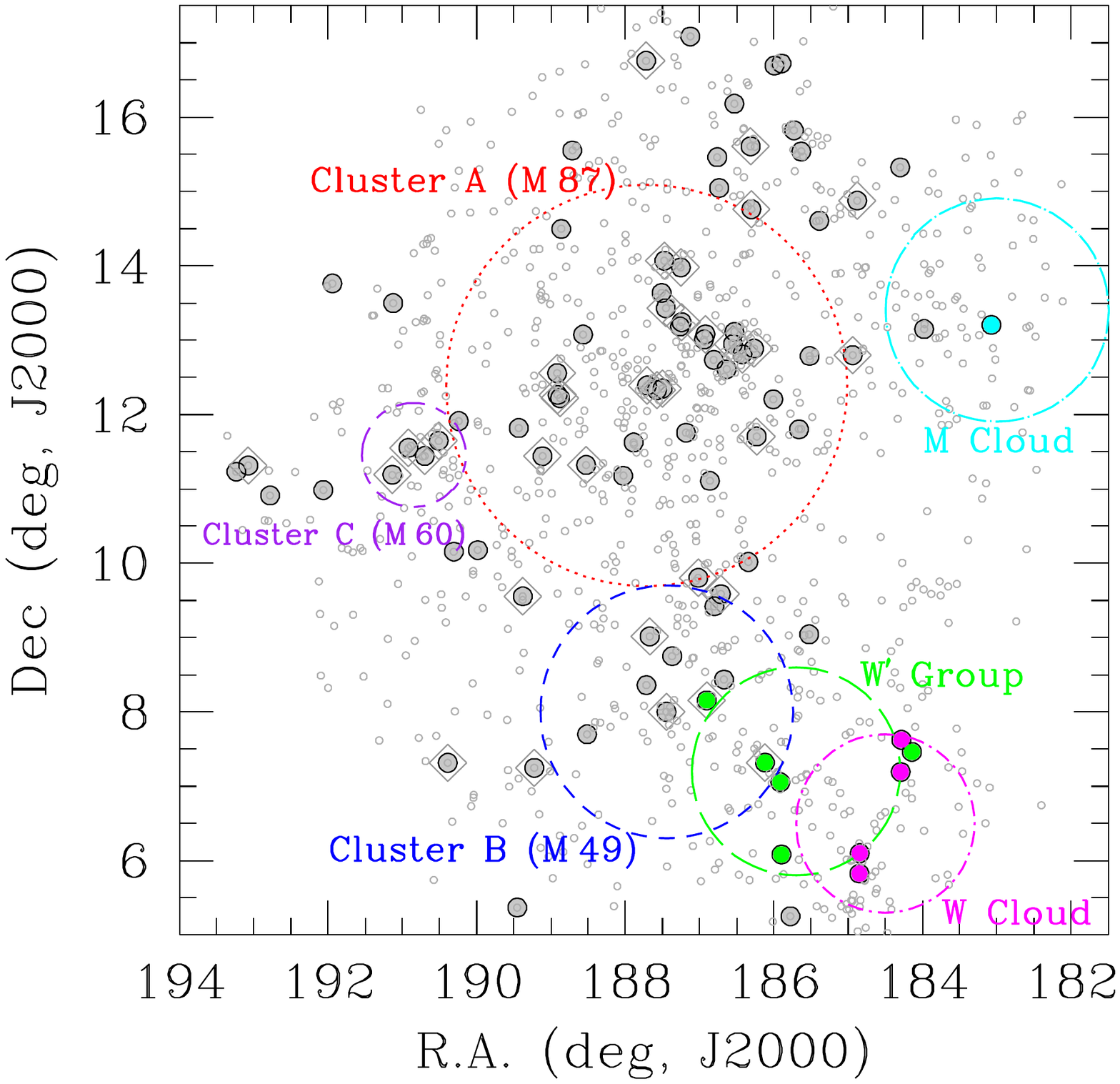}
\includegraphics[scale=0.4]{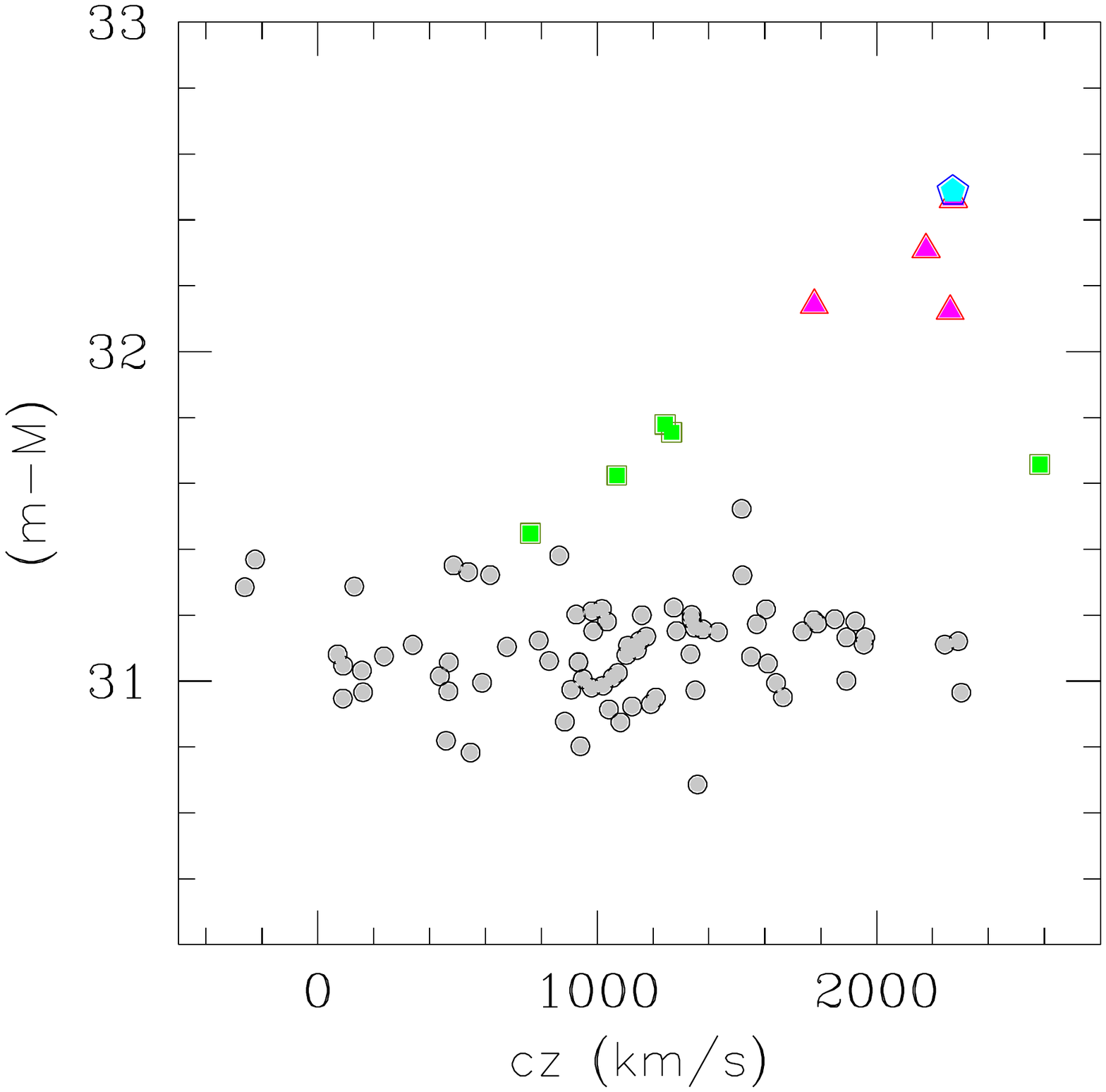}
\vspace{-2cm}
\caption{{Left: Map of the sky positions of the 89 galaxies in our
    sample, shown with large filled symbols. Small empty dots plot all
    possible and confirmed VCC members brighter than $B_T\approx18$
    mag. The 36 galaxies also in the ACSVCS are marked with gray
    diamonds. The substructures discussed in the text are identified
    by large symbols of various colors, using the locations and radii
    from \citet[][see text]{boselli14}. Note that these circles
    provide only rough indications of the structures; for instance,
    only the centermost galaxy in the ``M'' circle is at the distance
    of the M cloud, while we identify a total of 9 galaxies as
    belonging to either W or \Wprime\ based on their distances. The
    possible members of the \Wprime group and of the W and M clouds
    are also identified using the same color coding of the respective
    circles.  Right: Hubble diagram for the 89 galaxies in our
    sample. Symbol and color coding is the same as in left panel.}
\label{virgomap}}
\end{center}
\end{figure*}

\section{Results}
\label{sec_results}

\subsection{SBF distances for bright galaxies in the NGVS}
We now apply the color calibrations described in the previous sections
to derive the distance to each galaxy in our sample. We select the
$\overline{M}_i$ versus \uz\ relation as our preferred calibration because
it has the lowest rms scatter of all the single-color calibrations, and adding a
second color such as \gi\ does not further reduce the scatter. For
the six galaxies where no $u^*$-band magnitudes available, we adopt 
the calibration of $\overline{M}_i$ based on the \gi/\gz\ color combination,
as it has smaller scatter than either of the two single-color options
that do not include the \ustar~band. 

Adopting these calibrations, we derive the distances reported in
Table~\ref{tab_sbf} for the 89 galaxies in the present sample.  We
point out a few noteworthy results:
\begin{itemize}
\item
We find that VCC\,0049 (NGC\,4168), which has been classified as a likely member
of the M~cloud \citep[][]{ftaclas84,binggeli93},
is at a distance $d=31.4\pm1.9$~Mpc ($m{-}M\approx32.5$~mag), nearly twice as
far as the main body of the Virgo cluster. 
\item
The four galaxies VCC\,0220, VCC\,0222, VCC\,0341, and  VCC\,0345,
classified by \citet[][]{binggeli93} as members of the W~cloud,
are all significantly behind the main Virgo cluster and
have a mean $d$ of $28.7\pm1.1$~Mpc. Their mean heliocentric velocity
is 2120~\kms\ with a dispersion of $\approx200$~\kms. The brightest of these is
VCC\,0345, or NGC\,4261, which is also the only morphologically
classified giant elliptical; we take it to be the brightest group galaxy~(BGG).
\item As also found by the ACSVCS papers, VCC\,0731 (NGC\,4365) and
  VCC\,1025 (NGC\,4434) are members of the Virgo \Wprime\ group at
  $m{-}M\approx31.75$ ($d\approx22.5$ Mpc).  In addition, we find
  VCC\,0648 and VCC\,0657 are likely \Wprime\ group members based on
  their distances, velocities, and small angular separations
  ($1.2\deg$ and $0.3\deg$, respectively) from NGC\,4365, the
  \Wprime~BGG.
\item
 VCC\,0199 (NGC\,4224) has an estimated $d=21.5\pm1.3$~Mpc, which is
 consistent with being a member of the \wprime\ group. However, its
 velocity of 2584~\kms\ is much more similar to the mean for the
 W~cloud galaxies, as compared to the typical $v\approx1200$ \kms\ of
 the \Wprime\ members (Figure \ref{virgomap}, right panel).  Its
 angular position is somewhat closer to the mean W~cloud position, but
 because W is $\approx\,$25\% more distant than \Wprime, its physical
 separation would be smaller from NGC\,4365 if it were at the mean $d$
 of the \Wprime~group than it would be from NGC\,4261 if it were at
 the $d$ of the W~cloud.  We suggest that VCC\,0199 may be a
 high-velocity member of the \Wprime\ group. However, because it is a
 spiral with significant dust, our confidence in the SBF distance is
 much weaker than if it were an elliptical galaxy, and we consider the
 membership assignment uncertain.
\item The remaining galaxies, marked with ``V'' in Table~\ref{tab_sbf}, have
a mean $\langle{m{-}M}\rangle=31.090\pm0.013$~mag, or $d=16.5\pm0.1$
Mpc, which is the same mean Virgo distance from B09 quoted above.
This is not entirely trivial, since only 43\% (34 of 79) of the
V~sample was in the ACSVCS.  Our median estimated measurement error is
0.14~mag; thus, the dispersion in the distance moduli for the V~sample
mainly reflects measurement error.

\item We derive the intrinsic dispersion in distance modulus for the
  galaxies in the V sample, $\sigma_{\rm Vir}$, used for deriving the
  cluster depth, as follows. If we call the total observed scatters in
  Figure \ref{compacal_ngvs} (labeled ``rms'' in the figure)
  $\sigma_{\overline{i}-fit}$, which include the depth effect, and the
  scatters in Figure \ref{colfit4} $\sigma_{\overline{M}_i-fit}$ which
  do not contain the galaxy distance, then:

$\sigma_{\rm Vir} = \sqrt{ \sigma^2_{\overline{i}-fit} -
    \sigma^2_{\overline{M}_i-fit} - \sigma^2_{m{-}M}}$\\ where
  $\sigma_{m{-}M}$ is the median error in the assumed distance
  modulus, $\sim$0.07 mag for the ACSVCS distance moduli (Table
  \ref{tab_all}).

  Using the rms values for the \uz\ calibration, we obtain
  $\sigma_{\rm Vir}\sim 0.07$ mag, or $\sigma (d_{\rm Vir})\sim0.55$
  Mpc. The mean calculated from the four calibrations reported in
  Figures \ref{compacal}-\ref{colfit4} is $\sigma(d_{\rm
    Vir})=0.6\pm0.1$ Mpc, corresponding to a $\pm2\sigma$ depth of
  $2.4\pm0.4$ Mpc, a result consistent with \citet{mei07xiii}.
\item
The V~sample includes VCC\,1727 (NGC\,4579/M\,58) with
$\mM=31.52\pm0.15$~mag. This is the only putative V galaxy with
$\mM>31.4$~mag; given its center-east location projected within the
A~subcluster, it is not a candidate \Wprime\ member.  VCC\,1727 is a
spiral, and these tend to be either in front of, or behind, the Virgo
core \citep[e.g.,][]{kelson00}.  In any case, the weighted mean V
distance modulus is unchanged if this galaxy is excluded.
\end{itemize}

Figure \ref{virgoers} presents a histogram of the distance moduli
given in Table~\ref{tab_sbf}. The bins spanning Virgo proper, the
\Wprime\ group, and the W and M clouds are indicated. The distribution
of the galaxies on the sky is shown in Figure \ref{virgomap} (left
panel). In the figure, the 89 galaxies in our sample and the 36
targets in common with the ACSVCS are indicated with filled symbols
and gray empty diamonds, respectively. We also plot all possible and
confirmed VCC members brighter than $B_T\approx18.0$ mag, from
\citet{binggeli85}.  Some of the substructures in Virgo, i.e. the A, B
and C clusters (centered on M\,87 M\,49, and M\,60 respectively), the
\Wprime group, and the W and M clouds are also indicated using circles
of different colors and line-type adopting the positions and radii
given in \citet{boselli14}\footnote{With respect to \citet{boselli14},
  we have slightly changed the location of the circles representing
  the W cloud and the \Wprime group loci, as well as their radii, to
  better match the location of the galaxies}.  With the present data
set, we find no significant difference in the mean distances of the
galaxies within the A, B and C subclusters. To further highlight the
presence of substructures, also shown in Figure \ref{virgoers}, we
plot the Hubble diagram of the 89 galaxies in our sample in the right
panel of Figure \ref{virgomap}.

The results do not change substantially when the two-color calibration based on
\gi\ and \gz\ is used instead of our preferred \uz\ calibration
for all galaxies, rather than only for the six galaxies currently lacking
\ustar~data.  The mean distance modulus of Virgo, after excluding \Wprime, W,
and M galaxies, is virtually unchanged.  Moreover, for VCC\,0657, one of the new
\Wprime\ candidates, we find $m{-}M=31.70\pm0.37$, which is consistent with the
adopted value of $31.45\pm0.31$~mag, but closer to the \Wprime\ mean of
$\approx\,$31.75~mag.

\begin{figure*}
\begin{center}\includegraphics[scale=0.32]{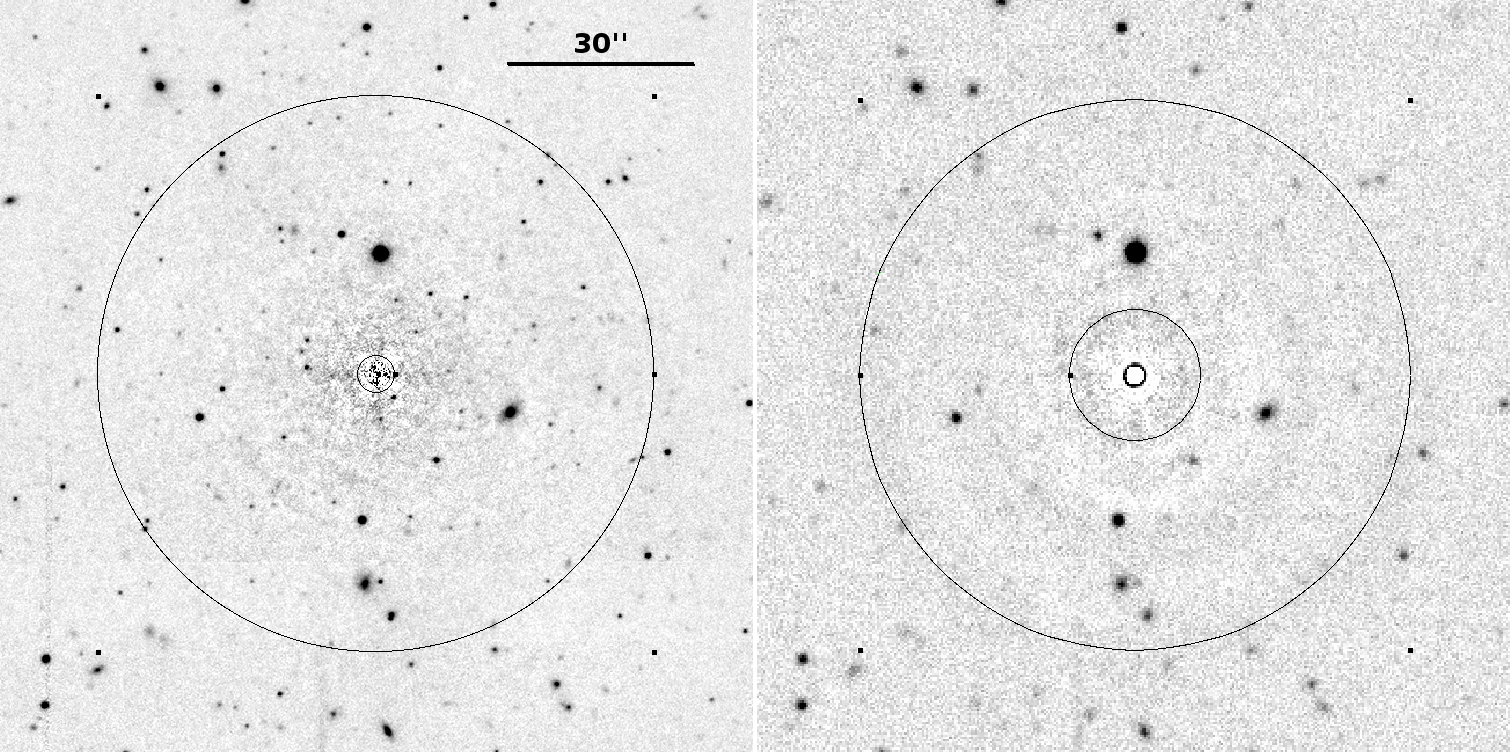}
\caption{Left: NGVS $i$-band residual image of VCC\,0648 (NGC\,4339).
  Right panel: $I$-band image of the same galaxy from the SBF survey
  of \citet{tonry97,tonry01}, obtained with the 2.4\,m Hiltner
  telescope at MDM Observatory in 1992. In both panels, the thin solid
  circles show the inner and outer radii of the annular region adopted
  to measure the amplitude of fluctuations.
\label{ngvs_t01}}
\end{center}
\end{figure*}

\begin{figure}
\includegraphics[scale=0.42]{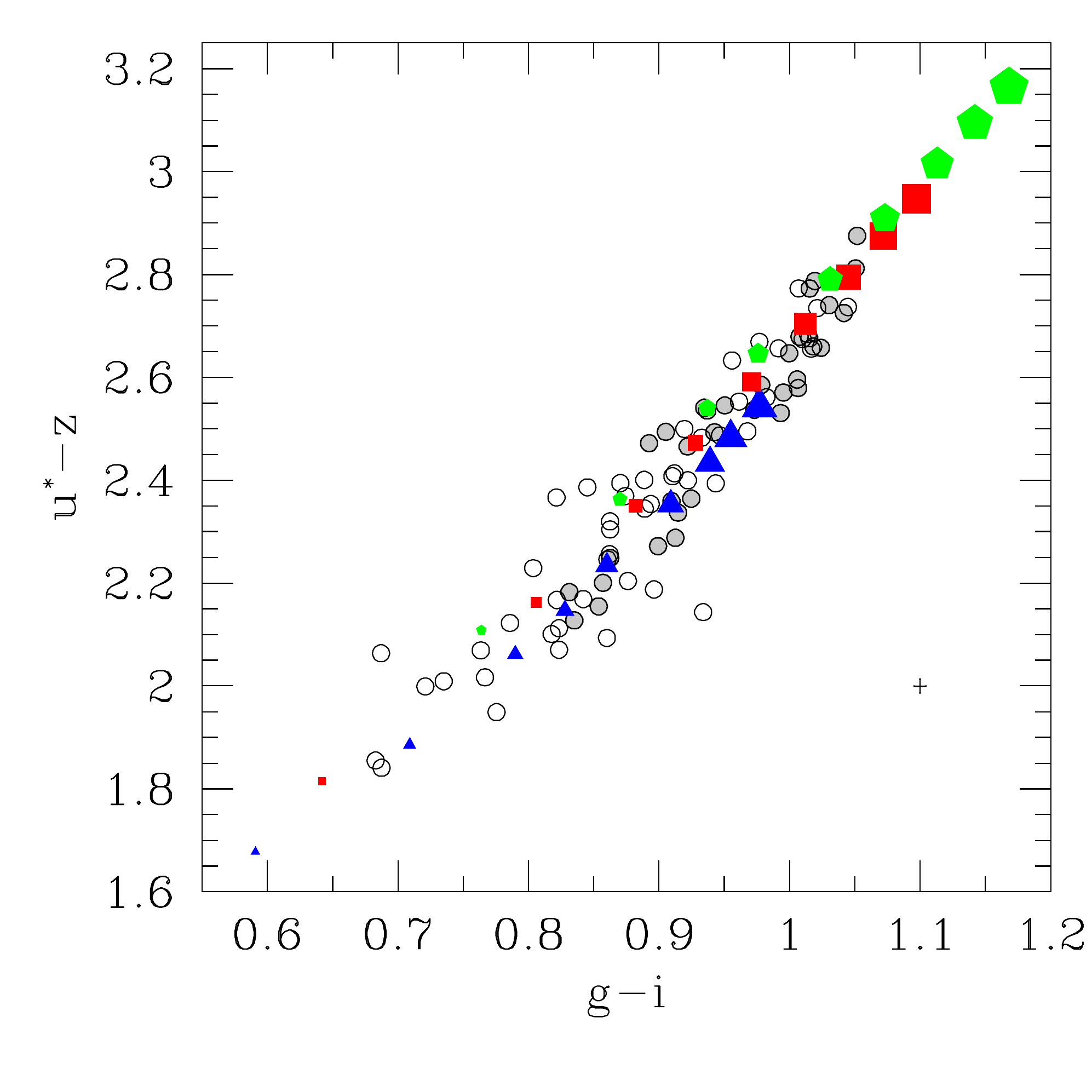}
\caption{Color-color diagram of NGVS galaxies presented in this work,
  compared to SPoT stellar population models. Empty circles mark the
  full sample of SBF galaxies, gray-filled circles mark the galaxies
  in common with the ACSVCS. The median errors are shown as a cross on
  the lower right side of the panel. SSP models for \feh$=-0.35,~0.0$
  and +0.4 are marked with blue triangles, red squares and green
  pentagons, respectively. The age range plotted is 1-14 Gyr (1, 2, 3,
  4, 6, 8, 10, 12, 14 Gyr) with increasing symbol size for older ages.
\label{colcol_models}}
\end{figure}

\begin{figure*}
\begin{center}
\includegraphics[scale=0.85,bb=0 295 560 600]{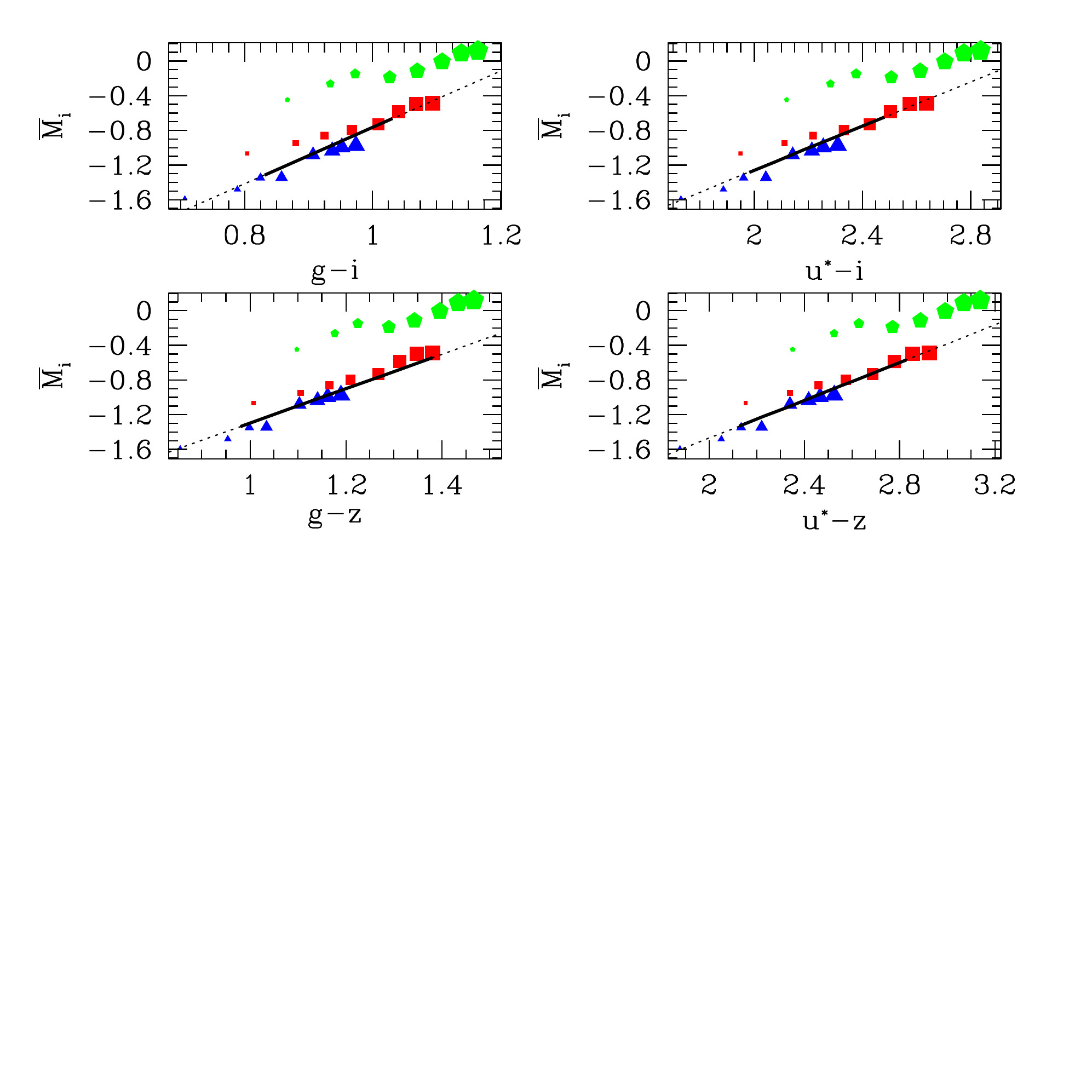}
\caption{SBF versus color calibrations: comparison of empirical
  relations and models. Symbols for models are the same as in Figure
  \ref{colcol_models}. In each panel the linear fit to the SBF versus
  color relation is marked with a black dotted line. The thick solid
  black line refers to the color interval of the objects used for
  deriving empirical relation.
\label{colfit_models}}
\end{center}
\end{figure*}

\subsection{Comparisons with literature and models}

We compared our distance results to those of other studies;
apart from our calibrating ACSVCS sample, the main overlap is with the
\citet[][hereafter Ton01]{tonry01} ground-based $I$-band SBF survey.
The present and Ton01 samples have 31 galaxies in common. 
The distances agree well on average, with
$(m{-}M)_\mathrm{NGVS}-(m{-}M)_\mathrm{Ton01}=-0.02\pm0.05$~mag and an rms of 
0.29~mag\footnote{The Ton01 distances have been revised
  using Eq.\,(A1) from \citet{blake10b} to correct for a small
  residual bias in Ton01 distances for galaxies with lowest SBF
  measurement quality.}. The rms is reduced to $\approx0.23$ mag when
the three most deviant galaxies are excluded from the comparison.
After rejecting the three outliers (discussed below),
the reduced $\chi^2$ of the differences for the remaining 28
galaxies is $\chi^2_\nu=1.18$.
The three largest outliers in the comparison to Ton01 are
VCC\,0648, VCC\,0958 and VCC\,1938 (NGC\,4339, NGC\,4419, and
NGC\,4638, respectively).  We discuss each of these in turn.

Our distance modulus for VCC\,0648/NGC\,4339 is $\approx0.65\pm0.20$
mag larger ($\approx$1.9 times more distant) than Ton01, or more than
3$\sigma$.  The Ton01 quality flags for the galaxy are below
average. Figure \ref{ngvs_t01} compares the NGVS $i$-band and Ton01
$I$-band (J.~Tonry, priv.~comm.)  images of the galaxy-subtracted
frame used for the respective SBF measurements. In the left panel we
show the NGVS image, which reveals a large number of sources, not
visible in the image used by Ton01 (right panel), many of which are
likely GCs in VCC\,0648.  The main difference between the images is
the seeing: the older data have a PSF FWHM of $\approx\,$1\farcs25,
while our NGVS image has FWHM $\approx\,$0\farcs55.  The poorer seeing
of the older data apparently caused, in this particular case, a poor
characterization of the GCLF because the data do not reach the
turnover magnitude.  As pointed out in Ton01 and other works, this can
lead to an underestimated $P_r$ correction, resulting in an
overestimated fluctuation amplitude and an artificially low distance
modulus.  In the case of the much deeper NGVS image, we see the
turnover in the GCLF. We conclude our distance is much more reliable
and, as discussed above, this galaxy is a newly recognized member of
the \Wprime\ group.

For VCC\,0958/NGC\,4419, Ton01 reported $(m{-}M)=30.65\pm0.25$~mag,
which places it in front of Virgo with a significance of 1.8$\sigma$.
The Ton01 quality flags are quite poor, as also indicated by the
relatively large quoted error. We find $(m{-}M)=31.29\pm0.18$~mag,
based on the \uz\ calibration. Using the two-color
\gi/\gz\ calibration, we derive $(m{-}M)=31.13 \pm 0.27$, which is
consistent with the \uz-calibrated distance, and closer to the Virgo
mean.  The Ton01 seeing for this galaxy was a relatively good
0\farcs85, but it is a highly inclined spiral with significant dust,
and therefore more difficult for the SBF method.  Other recent
distance estimates to the galaxy include $(m{-}M)=31.06\pm0.25$ from
the Tully-Fisher luminosity-linewidth relation by \citet[][revised to
  a mean Virgo distance modulus of $m{-}M=31.09$]{gavazzi99}, and
$(m{-}M)=31.24$ from \sna\ \citep{reindl05}. This \sna\ distance
modulus is based on the the peculiar supernova SN1984A and $H_0=60$
km/s/Mpc; assuming $H_0=70$ km/s/Mpc the distance modulus becomes
$(m{-}M)=30.91$ mag. The ``expected random error'' reported by
\citet{reindl05} is $\sigma_{m{-}M}=0.14$ mag, although the distance
moduli for single objects could be less accurate.  In summary, while
the \sna\ distances are not conclusive, based on our and Tully-Fisher
distances VCC\,0958 appears close to the mean distance of Virgo.

For VCC\,1938/NGC\,4638, we find $(m{-}M)=31.12\pm0.15$~mag, whereas
Ton01 reported $(m{-}M)=31.68\pm0.26$~mag, which differs by 1.9$\sigma$.
As shown in Table~\ref{tab_all}, this galaxy also has a measured ACSVCS distance
modulus of $31.19\pm0.07$~mag, which agrees with our value.
Note that the zero points of all three of these surveys are linked, but the
individual distances are independent.
Given our agreement with the much higher quality ACSVCS distance, we
again conclude that our measurement is likely accurate.

Finally, we have also done preliminary comparisons of our measured
SBF-color and color-color relations to predictions from the
Teramo-SPoT simple stellar population (SSP) models
\citep{cantiello03,raimondo05,raimondo09}\footnote{Models in
  $UBVRIJHK$ are available at the
  \url{www.oa-teramo.inaf.it/spot}. The new models used here, for SDSS
  and MegaCam photometric system, will be tabulated in next paper of
  this series.}.  The comparison in Figure \ref{colcol_models}
indicates that the range of observed colors corresponds with the color
sequence expected from the SSP models in the metallicity and age
ranges $-0.35 \leq\mathrm{[Fe/H]}\leq +0.4$ dex, $1 \leq t \leq
14$~Gyr. It is worth noting that the galaxies do not extend to the
region of the plot with the highest \feh\ values and oldest ages.

The four panels of Figure \ref{colfit_models} compare the linear fits
reported in Table~\ref{tab_1col} to SSP models predictions. In the
figure, the thick black lines represent the empirical fits and only
extend over the range of colors used in deriving them; dotted lines
indicate linear extrapolations to the fits.  The matching between
fitted relations and models with \feh between $-0.35$~dex and solar
(blue triangles and red squares in the figure), for ages older than
$\approx5$ Gyr is quite good.  The super-solar metallicity models (green
pentagons) predict fainter SBF magnitudes at the observed colors.
This appears to indicate that our SBF measurements are done in regions
of the galaxy with mean metallicity near solar, although it is also
true that the later stages of stellar evolution are less well
constrained at such high metallicities.  Very similar results from
model comparisons with ACS $I_{814}$ SBF data were reported by
\citet{blake10b}, and from spectroscopic analyses by
  \citet{trager00ii} and \citet{terlevich02}.

\section{Summary and Conclusions}
\label{sec_summa}

We have presented the first SBF measurements using the panoramic,
multi-band imaging data from the NGVS survey.  We described the
procedures for SBF analysis of the $i$-band data, and presented
multiple calibrations of the SBF absolute magnitude \Mbari\ in terms
of individual color indices and combinations of two different indices,
the so-called ``single-color'' and ``two-color'' calibrations.  By
comparing the measured scatter in the relations, we chose as our
preferred calibration the relation between \Mbari\ and \uz\ color.
For the range of colors covered by the present dataset, we find that
additional information from the \gi\ color measurements do not
significantly improve the rms scatter in the calibration.  For a small
number of galaxies that do not currently have \ustar~data, we instead
used an alternate dual-color calibration involving both \gi\ and \gz.
The observed scatter in both of these calibrations is
$\approx\,$0.11~mag, or about 5\% in distance, including both measurement
error and intrinsic scatter from stellar population effects.

In all, we have reported SBF distances for 89 galaxies brighter than
$B_T\approx13$ mag, the majority of which have had no previous SBF
measurements. The NGVS observation strategy was optimized for accurate
SBF measurements to galaxies at the mean distance of the Virgo
cluster, $\approx\,$16.5~Mpc, but the superb image quality of NGVS
enables high-quality SBF measurements out to $d>30$~Mpc.  For example,
we find five galaxies (VCC\,0049, VCC\,0220, VCC\,222, VCC\,0341 and
VCC\,0345) located 10-15~Mpc behind the main body of the Virgo
cluster; these are all likely members of either the M cloud at
$d\approx31$~Mpc or the W cloud at $\approx\,$28.7~Mpc.  The recession
velocities of these galaxies are also typically $\approx\,$1000
\kms\ larger than the mean recession velocity of Virgo.

The galaxies VCC\,0648 and VCC\,0657 are likely members of the Virgo
\Wprime\ group, based on having measured distances similar to the
value of $\approx\,$22.5~Mpc found for VCC\,0731 (NGC\,4365, the
brightest and most massive \Wprime\ galaxy) and small angular
separations from this galaxy.  VCC\,0199 may also be a member of the
\wprime\ group, based on its position and measured distance.  However,
its mean velocity is much closer to that of the W~cloud galaxies; we
therefore consider its membership uncertain. With the present data
set, we do not find any significant difference in the mean distances
to A, B and C subclusters, and obtain a cluster depth of $2.4\pm0.4$
Mpc (omitting the \Wprime\ group and more distant structures).

Comparisons of our color and SBF measurements with predictions from
SSP models are encouraging, both in supporting the adopted distance
zero point and suggesting interesting possibilities for future stellar
population studies using the full set of NGVS-SBF data.  Future papers
in this series will extend the catalog of SBF distances to fainter
$B_T$ magnitudes and bluer galaxies, where the single-color linear
calibration is expected to be a poorer representation of the
$\overline{M}_i$ versus color relation, and nonlinear calibrations,
possibly involving multiple colors, will likely be required. The
extended catalog will enable characterization of the 3-D structure of
the Virgo cluster and its various subcomponents to an unprecedented
level of detail.  Preliminary SBF distance moduli for three NGVS
galaxies several magnitudes fainter than the current sample limit have
already been presented by \citet{paudel17}.  These efforts are
ongoing, and we expect to produce an eventual catalog of $\approx250$
SBF-based distances for galaxies observed in the NGVS.

\acknowledgments This research has made use of the NASA/IPAC
Extragalactic Database (NED) which is operated by the Jet Propulsion
Laboratory, California Institute of Technology, under contract with
the National Aeronautics and Space Administration.  MC, GR and JPB
acknowledge support from the PRIN INAF-2014 `EXCALIBURS: EXtragalactic
distance scale CALIBration Using first-Rank Standard candles' project
(PI: G. Clementini). JPB gratefully acknowledges the hospitality of
Teramo Observatory. We thank John Tonry for the use of some software
and helpful consultations.

\facility{CFHT (MegaPrime/MegaCam)}

\clearpage
\bibliography{cantiello_sep17} 
\bibliographystyle{apj}


\clearpage
\begin{longtable*}{rrrrrcrc}
  \caption{Sample properties}\\
  \hline\noalign{\smallskip}\hline
  \multicolumn{1}{c}{VCC} &
  \multicolumn{1}{c}{RA (J2000)}&
  \multicolumn{1}{c}{Dec (J2000)}&
  \multicolumn{1}{c}{$B_T$} &
  \multicolumn{1}{c}{cz} &
  \multicolumn{1}{c}{$(m{-}M)_\mathrm{ACS}$} &
  \multicolumn{1}{c}{T$_{type}$} &
  \multicolumn{1}{c}{Alt. Name} \\
 \multicolumn{1}{c}{   } &
  \multicolumn{1}{c}{ (deg)   }&
  \multicolumn{1}{c}{ (deg)    }&
  \multicolumn{1}{c}{(mag)} &
  \multicolumn{1}{c}{$km/s$}&
  \multicolumn{1}{c}{ (mag)       } &
  \multicolumn{1}{c}{          } &
  \multicolumn{1}{c}{         } \\
\noalign{\smallskip}  \hline
   49 &   183.071933 &   13.205196 &   12.2 &    2273 &    \nodata            &     -4.90 $\pm$ 0.40 &  NGC\,4168 \\
  167 &   183.976608 &   13.149457 &   11.0 &     131 &    \nodata            &      3.00 $\pm$ 0.50 &  NGC\,4216 \\
  199 &   184.140781 &    7.462067 &   12.9 &    2584 &    \nodata            &      1.00 $\pm$ 0.30 &  NGC\,4224 \\
  220 &   184.282006 &    7.624276 &   13.0 &    2275 &    \nodata            &     -2.00 $\pm$ 0.50 &  NGC\,4233 \\
  222 &   184.291184 &    7.191575 &   12.7 &    2263 &    \nodata            &      1.00 $\pm$ 0.40 &  NGC\,4235 \\
  226 &   184.297607 &   15.324038 &   12.5 &     864 &    \nodata            &      4.00 $\pm$ 0.50 &  NGC\,4237 \\
  341 &   184.842724 &    6.098672 &   12.7 &    1777 &    \nodata            &      1.00 $\pm$ 0.50 &  NGC\,4260 \\
  345 &   184.846746 &    5.824897 &   11.3 &    2177 &    \nodata            &     -4.80 $\pm$ 0.40 &  NGC\,4261 \\
  355 &   184.877385 &   14.877653 &   12.4 &    1359 &    30.95 $\pm$   0.07 &     -2.60 $\pm$ 0.70 &  NGC\,4262 \\
  369 &   184.938669 &   12.798262 &   11.8 &    1021 &    31.00 $\pm$   0.07 &     -2.70 $\pm$ 0.70 &  NGC\,4267 \\
  483 &   185.386537 &   14.606137 &   12.1 &    1125 &    \nodata            &      5.10 $\pm$ 0.60 &  NGC\,4298 \\
  523 &   185.517183 &   12.787498 &   13.1 &    1520 &    \nodata            &     -2.00 $\pm$ 0.70 &  NGC\,4306 \\
  524 &   185.523566 &    9.043656 &   12.8 &    1055 &    \nodata            &      3.20 $\pm$ 0.70 &  NGC\,4307 \\
  559 &   185.630565 &   15.537915 &   12.6 &     158 &    \nodata            &      2.10 $\pm$ 0.90 &  NGC\,4312 \\
  570 &   185.660603 &   11.800900 &   12.7 &    1432 &    \nodata            &      2.10 $\pm$ 0.40 &  NGC\,4313 \\
  596 &   185.728792 &   15.822282 &   10.1 &    1571 &    \nodata            &      4.00 $\pm$ 0.30 &     NGC\,4321, M\,100 \\
  613 &   185.775720 &    5.250345 &   12.6 &    1665 &    \nodata            &     -0.80 $\pm$ 1.20 &              NGC\,4324 \\
  648 &   185.895606 &    6.081775 &   12.3 &    1266 &    \nodata            &     -4.60 $\pm$ 1.00 &              NGC\,4339 \\
  654 &   185.897017 &   16.722345 &   11.6 &     933 &    \nodata            &     -1.20 $\pm$ 0.60 &              NGC\,4340 \\
  657 &   185.912504 &    7.053998 &   12.6 &     761 &    \nodata            &     -3.20 $\pm$ 1.00 &              NGC\,4342 \\
  685 &   185.991010 &   16.693338 &   11.2 &    1210 &    \nodata            &     -1.80 $\pm$ 0.90 &              NGC\,4350 \\
  692 &   186.006313 &   12.204766 &   13.0 &    2303 &    \nodata            &      2.80 $\pm$ 1.70 &              NGC\,4351 \\
  731 &   186.117730 &    7.317770 &   10.5 &    1243 &    31.82 $\pm$   0.07 &     -4.80 $\pm$ 0.40 &              NGC\,4365 \\
  759 &   186.230973 &   11.704197 &   11.8 &     933 &    31.14 $\pm$   0.07 &     -1.30 $\pm$ 0.60 &              NGC\,4371 \\
  763 &   186.265603 &   12.886976 &   10.3 &    1017 &    31.34 $\pm$   0.07 &     -4.40 $\pm$ 1.20 &   NGC\,4374, M\,084 \\
  778 &   186.301410 &   14.762169 &   12.7 &    1338 &    31.24 $\pm$   0.07 &     -2.60 $\pm$ 0.60 &              NGC\,4377 \\
  784 &   186.311442 &   15.607421 &   12.7 &    1074 &    31.00 $\pm$   0.07 &     -2.70 $\pm$ 0.60 &              NGC\,4379 \\
  792 &   186.342391 &   10.016793 &   12.4 &     949 &    \nodata            &      2.40 $\pm$ 0.90 &              NGC\,4380 \\
  828 &   186.423673 &   12.810517 &   12.8 &     538 &    31.28 $\pm$   0.07 &     -4.80 $\pm$ 0.60 &              NGC\,4387 \\
  873 &   186.529766 &   13.111997 &   12.6 &     237 &    \nodata            &      3.20 $\pm$ 0.80 &              NGC\,4402 \\
  874 &   186.529812 &   16.180971 &   13.0 &    1735 &    \nodata            &      0.30 $\pm$ 1.10 &              NGC\,4405 \\
  881 &   186.548981 &   12.946240 &   10.1 &    -224 &    31.26 $\pm$   0.07 &     -4.80 $\pm$ 0.50 &   NGC\,4406, M\,086 \\
  912 &   186.634391 &   12.610708 &   12.4 &      91 &    \nodata            &      2.10 $\pm$ 1.10 &              NGC\,4407 \\
  929 &   186.668694 &    8.435697 &   13.1 &     907 &    \nodata            &     -0.90 $\pm$ 2.10 &              NGC\,4415 \\
  944 &   186.710867 &    9.584259 &   12.1 &     828 &    31.02 $\pm$   0.07 &     -1.90 $\pm$ 0.50 &              NGC\,4417 \\
  958 &   186.735107 &   15.047303 &   12.1 &    -261 &    \nodata            &      1.20 $\pm$ 0.90 &              NGC\,4419 \\
  966 &   186.760585 &   15.461466 &   12.4 &    1551 &    \nodata            &     -0.50 $\pm$ 0.80 &              NGC\,4421 \\
  979 &   186.798630 &    9.420773 &   12.3 &     437 &    \nodata            &      0.90 $\pm$ 0.50 &              NGC\,4424 \\
  984 &   186.805580 &   12.734765 &   12.3 &    1892 &    \nodata            &     -0.60 $\pm$ 1.20 &              NGC\,4425 \\
 1003 &   186.860486 &   11.107684 &   11.2 &    1104 &    \nodata            &     -0.80 $\pm$ 1.50 &              NGC\,4429 \\
 1025 &   186.902832 &    8.154342 &   13.1 &    1070 &    31.76 $\pm$   0.07 &     -4.70 $\pm$ 0.70 &              NGC\,4434 \\
 1030 &   186.918717 &   13.078984 &   11.8 &     791 &    31.11 $\pm$   0.07 &     -2.10 $\pm$ 0.50 &              NGC\,4435 \\
 1043 &   186.940202 &   13.008874 &   10.5 &      71 &    \nodata            &      0.60 $\pm$ 1.50 &              NGC\,4438 \\
 1062 &   187.016147 &    9.803712 &   11.4 &     547 &    30.93 $\pm$   0.07 &     -1.90 $\pm$ 0.40 &              NGC\,4442 \\
 1110 &   187.123295 &   17.085020 &   10.9 &    1954 &    \nodata            &      2.40 $\pm$ 0.70 &              NGC\,4450 \\
 1125 &   187.180447 &   11.755032 &   12.5 &     162 &    \nodata            &     -1.90 $\pm$ 0.80 &              NGC\,4452 \\
 1146 &   187.239835 &   13.241934 &   12.9 &     677 &    31.06 $\pm$   0.07 &     -4.90 $\pm$ 0.40 &              NGC\,4458 \\
 1154 &   187.250163 &   13.978475 &   11.4 &    1192 &    31.02 $\pm$   0.07 &     -1.60 $\pm$ 1.10 &              NGC\,4459 \\
 1158 &   187.262550 &   13.183801 &   11.5 &    1924 &    \nodata            &     -0.70 $\pm$ 1.30 &              NGC\,4461 \\
 1190 &   187.366793 &    8.749803 &   12.2 &     588 &    \nodata            &      0.20 $\pm$ 0.70 &              NGC\,4469 \\
 1226 &   187.444854 &    8.000490 &    9.3 &     981 &    31.12 $\pm$   0.07 &     -4.80 $\pm$ 0.50 &   NGC\,4472, M\,049 \\
 1231 &   187.453625 &   13.429436 &   11.1 &    2244 &    30.92 $\pm$   0.07 &     -4.70 $\pm$ 0.70 &              NGC\,4473 \\
 1242 &   187.473118 &   14.068584 &   12.6 &    1611 &    30.95 $\pm$   0.08 &     -1.90 $\pm$ 0.90 &              NGC\,4474 \\
 1250 &   187.496153 &   12.348719 &   12.9 &    1959 &    31.24 $\pm$   0.08 &     -2.90 $\pm$ 1.00 &              NGC\,4476 \\
 1253 &   187.509172 &   13.636533 &   11.3 &    1338 &    \nodata            &     -1.70 $\pm$ 0.70 &              NGC\,4477 \\
 1279 &   187.572571 &   12.328559 &   12.2 &    1349 &    31.16 $\pm$   0.07 &     -4.90 $\pm$ 0.40 &              NGC\,4478 \\
 1303 &   187.669345 &    9.015672 &   13.1 &     884 &    31.12 $\pm$   0.07 &     -1.40 $\pm$ 0.70 &              NGC\,4483 \\
 1316 &   187.705937 &   12.391122 &    9.6 &    1284 &    31.11 $\pm$   0.08 &     -4.30 $\pm$ 0.60 &   NGC\,4486, M\,087 \\
 1318 &   187.714059 &    8.360007 &   12.9 &     980 &    \nodata            &     -0.10 $\pm$ 0.60 &              NGC\,4488 \\
 1321 &   187.717714 &   16.758843 &   12.8 &     940 &    30.93 $\pm$   0.07 &     -4.70 $\pm$ 0.90 &              NGC\,4489 \\
 1368 &   187.885575 &   11.624746 &   12.7 &    1042 &    \nodata            &     -1.10 $\pm$ 1.10 &              NGC\,4497 \\
 1412 &   188.025974 &   11.176390 &   12.1 &    1334 &    \nodata            &     -1.70 $\pm$ 1.80 &              NGC\,4503 \\
 1535 &   188.512433 &    7.699310 &   10.0 &     617 &    \nodata            &     -1.90 $\pm$ 0.40 &              NGC\,4526 \\
 1537 &   188.525321 &   11.321253 &   12.7 &    1378 &    30.98 $\pm$   0.07 &     -2.00 $\pm$ 0.50 &              NGC\,4528 \\
 1552 &   188.566164 &   13.075314 &   12.6 &      90 &    \nodata            &     -0.10 $\pm$ 2.40 &              NGC\,4531 \\
 1588 &   188.711889 &   15.551642 &   12.5 &    1274 &    \nodata            &      6.20 $\pm$ 0.90 &              NGC\,4540 \\
 1615 &   188.860251 &   14.496350 &   11.0 &     486 &    \nodata            &      3.10 $\pm$ 0.50 &   NGC\,4548, M\,091 \\
 1619 &   188.877428 &   12.220753 &   12.5 &     459 &    30.93 $\pm$   0.07 &     -2.00 $\pm$ 0.70 &              NGC\,4550 \\
 1630 &   188.908131 &   12.263988 &   12.9 &    1176 &    31.05 $\pm$   0.07 &     -4.90 $\pm$ 0.40 &              NGC\,4551 \\
 1632 &   188.915884 &   12.556365 &   10.8 &     340 &    31.02 $\pm$   0.07 &     -4.60 $\pm$ 0.90 &   NGC\,4552, M\,089 \\
 1664 &   189.112434 &   11.439228 &   12.0 &    1142 &    31.01 $\pm$   0.07 &     -4.60 $\pm$ 0.70 &              NGC\,4564 \\
 1692 &   189.222419 &    7.246580 &   11.8 &    1787 &    31.17 $\pm$   0.07 &     -1.90 $\pm$ 1.00 &              NGC\,4570 \\
 1720 &   189.377334 &    9.555100 &   12.3 &    2292 &    31.07 $\pm$   0.07 &     -2.10 $\pm$ 0.60 &              NGC\,4578 \\
 1727 &   189.431389 &   11.818205 &   10.6 &    1517 &    \nodata            &      2.80 $\pm$ 0.60 &   NGC\,4579, M\,058 \\
 1730 &   189.451619 &    5.368521 &   12.6 &    1035 &    \nodata            &      1.60 $\pm$ 0.70 &              NGC\,4580 \\
 1813 &   189.983154 &   10.176147 &   11.5 &    1892 &    \nodata            &     -0.80 $\pm$ 0.80 &              NGC\,4596 \\
 1859 &   190.239811 &   11.912173 &   12.7 &    1640 &    \nodata            &      0.60 $\pm$ 1.40 &              NGC\,4606 \\
 1869 &   190.305385 &   10.155651 &   12.1 &    1850 &    \nodata            &     -1.70 $\pm$ 0.80 &              NGC\,4608 \\
 1883 &   190.386455 &    7.314879 &   12.6 &    1775 &    31.09 $\pm$   0.07 &     -2.00 $\pm$ 0.40 &              NGC\,4612 \\
 1903 &   190.509428 &   11.646945 &   10.8 &     467 &    30.86 $\pm$   0.07 &     -4.80 $\pm$ 0.40 &   NGC\,4621, M\,059 \\
 1938 &   190.697623 &   11.442507 &   12.1 &    1152 &    31.19 $\pm$   0.07 &     -2.60 $\pm$ 0.70 &              NGC\,4638 \\
 1978 &   190.916532 &   11.552700 &    9.8 &    1110 &    31.08 $\pm$   0.08 &     -4.60 $\pm$ 0.90 &   NGC\,4649, M\,060 \\
 1999 &   191.122489 &   13.498557 &   13.1 &     469 &    \nodata            &      0.00 $\pm$ 0.50 &              NGC\,4659 \\
 2000 &   191.133259 &   11.190501 &   11.9 &    1083 &    30.88 $\pm$   0.07 &     -4.60 $\pm$ 0.70 &              NGC\,4660 \\
 2058 &   191.939846 &   13.762808 &   11.6 &    1604 &    \nodata            &      4.70 $\pm$ 0.90 &              NGC\,4689 \\
 2066 &   192.063007 &   10.983284 &   12.2 &    1160 &    \nodata            &     -1.80 $\pm$ 0.70 &              NGC\,4694 \\
 2087 &   192.778237 &   10.912097 &   12.2 &     925 &    \nodata            &     -3.60 $\pm$ 1.30 &              NGC\,4733 \\
 2092 &   193.072900 &   11.313863 &   11.5 &    1351 &    31.03 $\pm$   0.07 &     -2.40 $\pm$ 1.10 &              NGC\,4754 \\
 2095 &   193.233227 &   11.230975 &   10.8 &     986 &    \nodata            &     -1.80 $\pm$ 0.90 &              NGC\,4762 \\
\hline
 \label{tab_all}
\end{longtable*}

\startlongtable
\begin{deluxetable}{rcrcrrrrrc}
\tabletypesize{\footnotesize}
\tablecaption{SBF and color measurements\label{tab_sbf}}
\tablehead{
  \colhead{VCC} &
  \colhead{Area} &
  \colhead{$\langle Rad\rangle$} &
  \colhead{\uz} &
  \colhead{\gi} &
  \colhead{\gz} &
  \colhead{$\overline{m_i}$} &
  \colhead{$(m{-}M)$\tablenotemark{1}} &
  \colhead{$D$} &
  \colhead{Comment\tablenotemark{2}} \\
  \colhead{   } &
  \colhead{(arcmin$^2$)} &
  \colhead{(arcsec)} &
  \colhead{(mag)} &
  \colhead{(mag)} &
  \colhead{(mag)} &
  \colhead{(mag)} &
  \colhead{(mag)} &
  \colhead{(Mpc)} &
  \colhead{     }
  } \startdata
 0049 &   4.11 &   51.5 &    2.623 $\pm$  0.007 &    0.983 $\pm$  0.002 &     1.214 $\pm$  0.007 &    31.63 $\pm$  0.09 &  32.49 $\pm$   0.13 &     31.4 $ \pm $    1.9 & M \\
 0167 &   1.05 &   59.2 &    2.501 $\pm$  0.030 &    0.999 $\pm$  0.013 &     1.234 $\pm$  0.029 &    30.30 $\pm$  0.14 &  31.29 $\pm$   0.23 &     18.1 $ \pm $    1.9 & V  \\
 0199 &   0.38 &   31.2 &    2.568 $\pm$  0.012 &    0.989 $\pm$  0.005 &     1.211 $\pm$  0.010 &    30.76 $\pm$  0.07 &  31.66 $\pm$   0.13 &     21.5 $ \pm $    1.3 & W/\Wprime? \\
 0220 &   1.02 &   32.9 &    2.737 $\pm$  0.008 &    1.033 $\pm$  0.004 &     1.302 $\pm$  0.007 &    31.75 $\pm$  0.12 &  32.46 $\pm$   0.15 &     31.1 $ \pm $    2.2 & W \\
 0222 &   0.37 &   28.0 &    2.546 $\pm$  0.015 &    0.965 $\pm$  0.006 &     1.209 $\pm$  0.012 &    31.21 $\pm$  0.15 &  32.12 $\pm$   0.19 &     26.6 $ \pm $    2.4 & W \\
 0226 &   0.64 &   54.5 &    2.171 $\pm$  0.042 &    0.816 $\pm$  0.012 &     1.069 $\pm$  0.041 &    30.04 $\pm$  0.08 &  31.38 $\pm$   0.23 &     18.9 $ \pm $    2.0 & V  \\
 0341 &   0.64 &   39.6 &    2.632 $\pm$  0.027 &    1.002 $\pm$  0.004 &     1.266 $\pm$  0.025 &    31.32 $\pm$  0.11 &  32.14 $\pm$   0.20 &     26.8 $ \pm $    2.5 & W \\
 0345 &   6.18 &   64.3 &    2.796 $\pm$  0.007 &    1.075 $\pm$  0.001 &     1.378 $\pm$  0.007 &    31.67 $\pm$  0.12 &  32.31 $\pm$   0.15 &     29.0 $ \pm $    2.0 & W \\
 0355 &   0.48 &   33.0 &    2.667 $\pm$  0.009 &    1.013 $\pm$  0.003 &     1.281 $\pm$  0.008 &    29.87 $\pm$  0.05 &  30.69 $\pm$   0.11 &     13.7 $ \pm $    0.7 & V  \\
 0369 &   2.96 &   45.4 &    2.815 $\pm$  0.005 &    1.106 $\pm$  0.007 &     1.346 $\pm$  0.008 &    30.32 $\pm$  0.04 &  30.99 $\pm$   0.09 &     15.7 $ \pm $    0.6 & V  \\
 0483 &   1.26 &   64.6 &    2.183 $\pm$  0.036 &    0.749 $\pm$  0.014 &     0.978 $\pm$  0.035 &    29.59 $\pm$  0.06 &  30.92 $\pm$   0.21 &     15.3 $ \pm $    1.5 & V  \\
 0523 &   1.82 &   32.2 &    2.308 $\pm$  0.010 &    0.895 $\pm$  0.005 &     1.076 $\pm$  0.010 &    30.11 $\pm$  0.06 &  31.32 $\pm$   0.12 &     18.4 $ \pm $    1.1 & V  \\
 0524 &   0.54 &   27.8 &    2.323 $\pm$  0.017 &    0.900 $\pm$  0.008 &     1.104 $\pm$  0.016 &    29.85 $\pm$  0.07 &  31.01 $\pm$   0.15 &     15.9 $ \pm $    1.1 & V  \\
 0559 &   1.25 &   36.4 &    2.194 $\pm$  0.015 &    0.866 $\pm$  0.005 &     1.050 $\pm$  0.014 &    29.72 $\pm$  0.09 &  31.03 $\pm$   0.16 &     16.1 $ \pm $    1.2 & V  \\
 0570 &   0.46 &   30.1 &    2.351 $\pm$  0.015 &    0.867 $\pm$  0.011 &     1.104 $\pm$  0.015 &    29.99 $\pm$  0.09 &  31.15 $\pm$   0.16 &     17.0 $ \pm $    1.2 & V  \\
 0596 &   4.11 &  179.0 &    2.087 $\pm$  0.036 &    0.767 $\pm$  0.012 &     1.052 $\pm$  0.036 &    29.75 $\pm$  0.06 &  31.17 $\pm$   0.21 &     17.2 $ \pm $    1.7 & V  \\
 0613 &   0.33 &   33.3 &    2.225 $\pm$  0.020 &    0.976 $\pm$  0.009 &     1.183 $\pm$  0.018 &    29.68 $\pm$  0.07 &  30.95 $\pm$   0.16 &     15.5 $ \pm $    1.2 & V  \\
 0648 &   1.01 &   28.9 &    2.764 $\pm$  0.004 &    1.061 $\pm$  0.001 &     1.320 $\pm$  0.004 &    31.07 $\pm$  0.06 &  31.75 $\pm$   0.09 &     22.4 $ \pm $    0.9 & \wprime\\
 0654 &   3.37 &   54.7 &    \nodata            &    1.020 $\pm$  0.003 &     1.281 $\pm$  0.010 &    30.25 $\pm$  0.04 &  31.06 $\pm$   0.14 &     16.3 $ \pm $    1.1 & V  \\
 0657 &   0.35 &   27.1 &    2.435 $\pm$  0.072 &    0.857 $\pm$  0.017 &     1.070 $\pm$  0.069 &    30.41 $\pm$  0.12 &  31.45 $\pm$   0.31 &     19.5 $ \pm $    2.7 & \wprime \\
 0685 &   3.67 &   55.3 &    \nodata            &    0.983 $\pm$  0.005 &     1.230 $\pm$  0.020 &    30.02 $\pm$  0.05 &  30.95 $\pm$   0.20 &     15.5 $ \pm $    1.4 & V  \\
 0692 &   0.54 &   43.0 &    1.936 $\pm$  0.040 &    0.737 $\pm$  0.030 &     0.866 $\pm$  0.040 &    29.37 $\pm$  0.07 &  30.96 $\pm$   0.22 &     15.6 $ \pm $    1.6 & V  \\
 0731 &   3.96 &   52.2 &    2.811 $\pm$  0.000 &    1.067 $\pm$  0.000 &     1.311 $\pm$  0.000 &    31.15 $\pm$  0.06 &  31.78 $\pm$   0.06 &     22.7 $ \pm $    0.6 & \wprime \\
 0759 &   0.83 &   54.3 &    2.717 $\pm$  0.008 &    1.069 $\pm$  0.006 &     1.340 $\pm$  0.007 &    30.30 $\pm$  0.07 &  31.06 $\pm$   0.12 &     16.3 $ \pm $    0.9 & V  \\
 0763 &   5.91 &   61.5 &    2.866 $\pm$  0.002 &    1.114 $\pm$  0.001 &     1.371 $\pm$  0.001 &    30.61 $\pm$  0.04 &  31.22 $\pm$   0.06 &     17.5 $ \pm $    0.5 & V  \\
 0778 &   3.26 &   48.2 &    2.624 $\pm$  0.016 &    1.010 $\pm$  0.015 &     1.240 $\pm$  0.021 &    30.33 $\pm$  0.05 &  31.19 $\pm$   0.14 &     17.3 $ \pm $    1.1 & V  \\
 0784 &   3.22 &   46.0 &    2.650 $\pm$  0.011 &    1.036 $\pm$  0.004 &     1.259 $\pm$  0.011 &    30.22 $\pm$  0.04 &  31.02 $\pm$   0.12 &     16.0 $ \pm $    0.9 & V  \\
 0792 &   0.41 &   54.1 &    2.480 $\pm$  0.017 &    0.963 $\pm$  0.008 &     1.161 $\pm$  0.016 &    30.01 $\pm$  0.06 &  31.01 $\pm$   0.15 &     15.9 $ \pm $    1.1 & V  \\
 0828 &   1.20 &   33.3 &    2.639 $\pm$  0.012 &    1.049 $\pm$  0.006 &     1.261 $\pm$  0.012 &    30.50 $\pm$  0.08 &  31.33 $\pm$   0.14 &     18.5 $ \pm $    1.2 & V  \\
 0873 &   0.52 &   34.4 &    2.168 $\pm$  0.016 &    0.874 $\pm$  0.007 &     1.039 $\pm$  0.015 &    29.73 $\pm$  0.05 &  31.07 $\pm$   0.14 &     16.4 $ \pm $    1.1 & V  \\
 0874 &   0.39 &   32.4 &    2.205 $\pm$  0.011 &    0.829 $\pm$  0.005 &     1.030 $\pm$  0.011 &    29.86 $\pm$  0.07 &  31.15 $\pm$   0.13 &     17.0 $ \pm $    1.0 & V  \\
 0881 &   2.82 &   46.3 &    2.777 $\pm$  0.001 &    1.058 $\pm$  0.001 &     1.335 $\pm$  0.001 &    30.69 $\pm$  0.10 &  31.37 $\pm$   0.10 &     18.8 $ \pm $    0.9 & V  \\
 0912 &   0.96 &   52.5 &    2.278 $\pm$  0.037 &    0.898 $\pm$  0.017 &     1.063 $\pm$  0.036 &    29.82 $\pm$  0.08 &  31.05 $\pm$   0.22 &     16.2 $ \pm $    1.6 & V  \\
 0929 &   2.67 &   39.2 &    2.469 $\pm$  0.013 &    0.909 $\pm$  0.004 &     1.149 $\pm$  0.012 &    29.97 $\pm$  0.04 &  30.97 $\pm$   0.12 &     15.7 $ \pm $    0.9 & V  \\
 0944 &   2.11 &   49.5 &    2.371 $\pm$  0.021 &    0.956 $\pm$  0.007 &     1.181 $\pm$  0.020 &    29.95 $\pm$  0.11 &  31.06 $\pm$   0.19 &     16.3 $ \pm $    1.4 & V  \\
 0958 &   0.85 &   40.8 &    2.205 $\pm$  0.025 &    0.918 $\pm$  0.010 &     1.139 $\pm$  0.024 &    29.98 $\pm$  0.07 &  31.29 $\pm$   0.18 &     18.1 $ \pm $    1.5 & V  \\
 0966 &   1.78 &   38.7 &    \nodata            &    0.947 $\pm$  0.003 &     1.229 $\pm$  0.005 &    30.09 $\pm$  0.06 &  31.07 $\pm$   0.13 &     16.4 $ \pm $    1.0 & V  \\
 0979 &   0.25 &   43.3 &    2.019 $\pm$  0.020 &    0.812 $\pm$  0.007 &     0.970 $\pm$  0.020 &    29.53 $\pm$  0.10 &  31.02 $\pm$   0.18 &     16.0 $ \pm $    1.3 & V  \\
 0984 &   3.04 &   53.7 &    2.285 $\pm$  0.046 &    0.946 $\pm$  0.018 &     1.123 $\pm$  0.045 &    29.79 $\pm$  0.07 &  31.00 $\pm$   0.24 &     15.9 $ \pm $    1.7 & V  \\
 1003 &   4.68 &   61.1 &    2.766 $\pm$  0.004 &    1.074 $\pm$  0.002 &     1.388 $\pm$  0.003 &    30.38 $\pm$  0.04 &  31.08 $\pm$   0.07 &     16.4 $ \pm $    0.6 & V  \\
 1025 &   2.42 &   38.1 &    2.612 $\pm$  0.009 &    0.976 $\pm$  0.003 &     1.218 $\pm$  0.009 &    30.78 $\pm$  0.04 &  31.62 $\pm$   0.11 &     21.1 $ \pm $    1.1 & \wprime \\
 1030 &   1.21 &   37.5 &    2.461 $\pm$  0.005 &    0.975 $\pm$  0.002 &     1.172 $\pm$  0.005 &    30.10 $\pm$  0.06 &  31.12 $\pm$   0.10 &     16.8 $ \pm $    0.8 & V  \\
 1043 &   1.84 &   77.5 &    2.398 $\pm$  0.024 &    0.911 $\pm$  0.011 &     1.231 $\pm$  0.023 &    29.99 $\pm$  0.07 &  31.08 $\pm$   0.18 &     16.5 $ \pm $    1.3 & V  \\
 1062 &   2.21 &   46.6 &    2.654 $\pm$  0.005 &    1.045 $\pm$  0.002 &     1.305 $\pm$  0.005 &    29.98 $\pm$  0.03 &  30.78 $\pm$   0.08 &     14.3 $ \pm $    0.5 & V  \\
 1110 &   1.62 &   80.4 &    2.508 $\pm$  0.016 &    0.961 $\pm$  0.004 &     1.244 $\pm$  0.015 &    30.14 $\pm$  0.07 &  31.11 $\pm$   0.15 &     16.7 $ \pm $    1.1 & V  \\
 1125 &   1.44 &   36.5 &    2.305 $\pm$  0.039 &    0.928 $\pm$  0.006 &     1.091 $\pm$  0.039 &    29.77 $\pm$  0.11 &  30.97 $\pm$   0.23 &     15.6 $ \pm $    1.7 & V  \\
 1146 &   4.91 &   53.0 &    2.415 $\pm$  0.011 &    0.955 $\pm$  0.004 &     1.172 $\pm$  0.010 &    30.04 $\pm$  0.05 &  31.10 $\pm$   0.12 &     16.6 $ \pm $    0.9 & V  \\
 1154 &   2.29 &   38.3 &    2.925 $\pm$  0.001 &    1.094 $\pm$  0.001 &     1.442 $\pm$  0.001 &    30.38 $\pm$  0.06 &  30.93 $\pm$   0.08 &     15.3 $ \pm $    0.5 & V  \\
 1158 &   3.07 &   42.1 &    2.746 $\pm$  0.004 &    1.017 $\pm$  0.002 &     1.256 $\pm$  0.004 &    30.48 $\pm$  0.11 &  31.18 $\pm$   0.13 &     17.2 $ \pm $    1.0 & V  \\
 1190 &   0.76 &   38.6 &    2.563 $\pm$  0.014 &    1.003 $\pm$  0.005 &     1.239 $\pm$  0.014 &    30.10 $\pm$  0.09 &  31.00 $\pm$   0.16 &     15.8 $ \pm $    1.1 & V  \\
 1226 &  15.51 &   98.0 &    2.863 $\pm$  0.002 &    1.058 $\pm$  0.001 &     1.369 $\pm$  0.002 &    30.64 $\pm$  0.07 &  31.21 $\pm$   0.09 &     17.5 $ \pm $    0.7 & V  \\
 1231 &   5.11 &   57.2 &    2.680 $\pm$  0.004 &    1.027 $\pm$  0.002 &     1.302 $\pm$  0.004 &    30.33 $\pm$  0.03 &  31.11 $\pm$   0.07 &     16.7 $ \pm $    0.6 & V  \\
 1242 &   3.61 &   55.7 &    2.322 $\pm$  0.019 &    0.904 $\pm$  0.009 &     1.118 $\pm$  0.018 &    29.86 $\pm$  0.05 &  31.05 $\pm$   0.15 &     16.2 $ \pm $    1.1 & V  \\
 1250 &   0.61 &   22.2 &    2.291 $\pm$  0.013 &    0.904 $\pm$  0.002 &     1.124 $\pm$  0.013 &    29.93 $\pm$  0.07 &  31.13 $\pm$   0.14 &     16.9 $ \pm $    1.1 & V  \\
 1253 &   3.66 &   54.0 &    2.841 $\pm$  0.003 &    1.076 $\pm$  0.002 &     1.368 $\pm$  0.003 &    30.59 $\pm$  0.08 &  31.20 $\pm$   0.10 &     17.4 $ \pm $    0.8 & V  \\
 1279 &   0.51 &   24.6 &    2.615 $\pm$  0.003 &    1.014 $\pm$  0.001 &     1.310 $\pm$  0.003 &    30.32 $\pm$  0.09 &  31.16 $\pm$   0.10 &     17.1 $ \pm $    0.8 & V  \\
 1303 &   1.31 &   37.2 &    2.560 $\pm$  0.022 &    0.939 $\pm$  0.006 &     1.217 $\pm$  0.021 &    29.98 $\pm$  0.05 &  30.88 $\pm$   0.16 &     15.0 $ \pm $    1.1 & V  \\
 1316 &   9.13 &   82.8 &    2.890 $\pm$  0.001 &    1.091 $\pm$  0.001 &     1.529 $\pm$  0.000 &    30.61 $\pm$  0.03 &  31.15 $\pm$   0.04 &     17.0 $ \pm $    0.4 & V  \\
 1318 &   1.52 &   33.9 &    2.438 $\pm$  0.009 &    0.910 $\pm$  0.003 &     1.140 $\pm$  0.008 &    29.95 $\pm$  0.04 &  30.98 $\pm$   0.10 &     15.7 $ \pm $    0.7 & V  \\
 1321 &   2.44 &   37.8 &    \nodata            &    0.946 $\pm$  0.003 &     1.171 $\pm$  0.009 &    29.75 $\pm$  0.05 &  30.80 $\pm$   0.14 &     14.5 $ \pm $    0.9 & V  \\
 1368 &   2.96 &   47.6 &    2.399 $\pm$  0.018 &    0.936 $\pm$  0.008 &     1.175 $\pm$  0.017 &    29.80 $\pm$  0.04 &  30.91 $\pm$   0.15 &     15.2 $ \pm $    1.0 & V  \\
 1412 &   1.07 &   38.7 &    2.845 $\pm$  0.007 &    1.098 $\pm$  0.002 &     1.350 $\pm$  0.007 &    30.44 $\pm$  0.05 &  31.08 $\pm$   0.10 &     16.5 $ \pm $    0.8 & V  \\
 1535 &   1.57 &   54.4 &    2.637 $\pm$  0.005 &    1.021 $\pm$  0.002 &     1.308 $\pm$  0.005 &    30.50 $\pm$  0.09 &  31.32 $\pm$   0.12 &     18.4 $ \pm $    1.0 & V  \\
 1537 &   2.70 &   44.3 &    2.403 $\pm$  0.032 &    0.942 $\pm$  0.009 &     1.154 $\pm$  0.031 &    30.04 $\pm$  0.08 &  31.16 $\pm$   0.20 &     17.0 $ \pm $    1.6 & V  \\
 1552 &   2.20 &   45.6 &    2.492 $\pm$  0.007 &    0.965 $\pm$  0.003 &     1.194 $\pm$  0.007 &    29.93 $\pm$  0.29 &  30.95 $\pm$   0.31 &     15.5 $ \pm $    2.2 & V  \\
 1588 &   0.50 &   47.7 &    2.133 $\pm$  0.024 &    0.827 $\pm$  0.006 &     1.018 $\pm$  0.024 &    29.83 $\pm$  0.03 &  31.22 $\pm$   0.17 &     17.6 $ \pm $    1.4 & V  \\
 1615 &   2.64 &  129.3 &    2.514 $\pm$  0.024 &    0.911 $\pm$  0.012 &     1.248 $\pm$  0.023 &    30.37 $\pm$  0.07 &  31.35 $\pm$   0.18 &     18.6 $ \pm $    1.5 & V  \\
 1619 &   2.06 &   40.7 &    2.291 $\pm$  0.016 &    0.924 $\pm$  0.007 &     1.104 $\pm$  0.015 &    29.59 $\pm$  0.06 &  30.82 $\pm$   0.14 &     14.6 $ \pm $    1.0 & V  \\
 1630 &   2.43 &   39.4 &    2.792 $\pm$  0.008 &    1.086 $\pm$  0.004 &     1.314 $\pm$  0.008 &    30.45 $\pm$  0.05 &  31.14 $\pm$   0.11 &     16.9 $ \pm $    0.8 & V  \\
 1632 &   4.07 &   51.3 &    2.822 $\pm$  0.002 &    1.085 $\pm$  0.001 &     1.383 $\pm$  0.002 &    30.46 $\pm$  0.08 &  31.11 $\pm$   0.10 &     16.7 $ \pm $    0.7 & V  \\
 1664 &   3.81 &   54.0 &    2.579 $\pm$  0.012 &    0.981 $\pm$  0.004 &     1.251 $\pm$  0.012 &    30.19 $\pm$  0.07 &  31.09 $\pm$   0.14 &     16.6 $ \pm $    1.0 & V  \\
 1692 &   3.20 &   61.5 &    2.345 $\pm$  0.044 &    0.937 $\pm$  0.012 &     1.204 $\pm$  0.041 &    30.04 $\pm$  0.07 &  31.18 $\pm$   0.23 &     17.2 $ \pm $    1.8 & V  \\
 1720 &   5.56 &   56.3 &    2.609 $\pm$  0.008 &    0.970 $\pm$  0.003 &     1.232 $\pm$  0.008 &    30.28 $\pm$  0.07 &  31.12 $\pm$   0.12 &     16.8 $ \pm $    0.9 & V  \\
 1727 &   1.24 &  114.7 &    2.457 $\pm$  0.012 &    0.934 $\pm$  0.005 &     1.221 $\pm$  0.011 &    30.47 $\pm$  0.09 &  31.52 $\pm$   0.15 &     20.2 $ \pm $    1.4 & V  \\
 1730 &   2.71 &   57.8 &    2.435 $\pm$  0.018 &    0.936 $\pm$  0.009 &     1.203 $\pm$  0.017 &    30.14 $\pm$  0.05 &  31.18 $\pm$   0.15 &     17.2 $ \pm $    1.2 & V  \\
 1813 &   7.66 &   74.8 &    2.705 $\pm$  0.007 &    0.993 $\pm$  0.003 &     1.256 $\pm$  0.006 &    30.39 $\pm$  0.09 &  31.13 $\pm$   0.12 &     16.9 $ \pm $    1.0 & V  \\
 1859 &   0.68 &   39.3 &    2.116 $\pm$  0.021 &    0.791 $\pm$  0.006 &     0.981 $\pm$  0.021 &    29.59 $\pm$  0.08 &  30.99 $\pm$   0.17 &     15.8 $ \pm $    1.3 & V  \\
 1869 &   1.76 &   54.7 &    2.832 $\pm$  0.004 &    1.037 $\pm$  0.001 &     1.297 $\pm$  0.003 &    30.59 $\pm$  0.08 &  31.19 $\pm$   0.10 &     17.3 $ \pm $    0.8 & V  \\
 1883 &   0.80 &   23.7 &    2.444 $\pm$  0.002 &    0.953 $\pm$  0.001 &     1.182 $\pm$  0.002 &    30.15 $\pm$  0.12 &  31.18 $\pm$   0.13 &     17.3 $ \pm $    1.0 & V  \\
 1903 &   2.63 &   49.3 &    2.759 $\pm$  0.002 &    1.058 $\pm$  0.001 &     1.353 $\pm$  0.002 &    30.26 $\pm$  0.04 &  30.97 $\pm$   0.07 &     15.6 $ \pm $    0.5 & V  \\
 1938 &   3.08 &   55.8 &    2.558 $\pm$  0.019 &    0.937 $\pm$  0.006 &     1.226 $\pm$  0.019 &    30.21 $\pm$  0.05 &  31.12 $\pm$   0.15 &     16.8 $ \pm $    1.2 & V  \\
 1978 &   3.04 &   47.2 &    2.964 $\pm$  0.001 &    1.098 $\pm$  0.000 &     1.427 $\pm$  0.001 &    30.64 $\pm$  0.07 &  31.11 $\pm$   0.08 &     16.7 $ \pm $    0.6 & V  \\
 1999 &   1.00 &   26.3 &    2.487 $\pm$  0.007 &    0.934 $\pm$  0.002 &     1.195 $\pm$  0.007 &    30.07 $\pm$  0.08 &  31.06 $\pm$   0.12 &     16.3 $ \pm $    0.9 & V  \\
 2000 &   3.57 &   55.1 &    2.242 $\pm$  0.023 &    0.895 $\pm$  0.007 &     1.060 $\pm$  0.022 &    29.61 $\pm$  0.04 &  30.88 $\pm$   0.16 &     15.0 $ \pm $    1.1 & V  \\
 2058 &   2.08 &   88.2 &    2.189 $\pm$  0.027 &    0.863 $\pm$  0.010 &     1.109 $\pm$  0.026 &    29.91 $\pm$  0.07 &  31.22 $\pm$   0.19 &     17.5 $ \pm $    1.5 & V  \\
 2066 &   1.65 &   63.0 &    1.986 $\pm$  0.037 &    0.751 $\pm$  0.017 &     0.911 $\pm$  0.037 &    29.64 $\pm$  0.07 &  31.20 $\pm$   0.21 &     17.4 $ \pm $    1.7 & V  \\
 2087 &   3.71 &   49.4 &    2.473 $\pm$  0.015 &    0.944 $\pm$  0.004 &     1.179 $\pm$  0.014 &    30.21 $\pm$  0.06 &  31.20 $\pm$   0.14 &     17.4 $ \pm $    1.1 & V  \\
 2092 &   3.67 &   55.0 &    \nodata            &    1.053 $\pm$  0.002 &     1.325 $\pm$  0.007 &    30.26 $\pm$  0.05 &  30.97 $\pm$   0.13 &     15.6 $ \pm $    0.9 & V  \\
 2095 &   2.63 &   53.3 &    \nodata            &    0.874 $\pm$  0.005 &     1.223 $\pm$  0.015 &    30.02 $\pm$  0.06 &  31.15 $\pm$   0.18 &     17.0 $ \pm $    1.4 & V  \\
\enddata
\tablenotetext{1}{The preferred distance modulus is derived adopting the \Mbari versus \uz\ calibration (Table~\ref{tab_1col}) when $u^*$-band is available, otherwise it is derived from the two-color calibration \Mbari versus \gi/\gz\  (Table~\ref{tab_2col}).}
\tablenotetext{2}{Labels used: V: Virgo cluster proper (A or B clouds); \wprime: Virgo \wprime\ member or candidate; M or W: member of either the $\sim$ twice more distant M or W clouds.}
\end{deluxetable}
\setcounter{table}{2}
\begin{deluxetable}{cccc}
\tablecaption{Coefficients of the fits $\Mbari=\alpha+\beta\times[color-refcolor]$\label{tab_1col}} 
\tablewidth{0pt}
 \tablehead{
\colhead{Color} & \colhead{$\alpha$ (mag)} & \colhead{$\beta$} & \colhead{$refcolor$} 
}
\startdata
\gi   & -0.93 $\pm$  0.04 & 3.25 $\pm$  0.42 & 0.95 \\
\gz   & -0.90 $\pm$  0.04 & 1.98 $\pm$  0.16 & 1.20  \\
\ui   & -0.88 $\pm$  0.04 & 1.27 $\pm$  0.06 & 2.30  \\
\uz   & -0.93 $\pm$  0.04 & 1.09 $\pm$  0.04 & 2.50  \\
\enddata
\end{deluxetable}

\begin{deluxetable}{ccccc}
  \tablecaption{Coefficients of the fits $\Mbari=A+B\times color_1+C\times color_2$\label{tab_2col}} 
\tablewidth{0pt}
 \tablehead{
\colhead{color$_1$} & \colhead{color$_2$} & \colhead{$A$ (mag)} & \colhead{$B$} & \colhead{$C$} 
}
\startdata
\ui & \gz  & -3.78 & +0.77 & +0.94  \\
\ug & \gi  & -3.98 & +0.71 & +2.22   \\
\uz & \gi  & -3.78 & +0.86 & +0.75   \\
\gi & \gz  & -3.83 & +1.98 & +0.86   \\
\enddata
\end{deluxetable}

\end{document}